\documentclass[sts, preprint, twocolumn]{imsart}

\RequirePackage[OT1]{fontenc}
\usepackage{amsthm,amsmath}
\usepackage[authoryear]{natbib}
\usepackage[]{graphicx}
\usepackage{epstopdf}

\graphicspath{{new_images/}}

\usepackage{framed}
\usepackage{enumitem}
\usepackage{bm}
\usepackage{amssymb}

\begin{document}

\begin{frontmatter}

% "Title of the paper"
\title{A comparison of inferential methods for highly non-linear state space models in ecology and epidemiology}
\runtitle{Inference for highly non-linear dynamical models}

% indicate corresponding author with \corref{}
% \author{\fnms{John} \snm{Smith}\corref{}\ead[label=e1]{smith@foo.com}\thanksref{t1}}
% \thankstext{t1}{Thanks to somebody} 
% \address{line 1\\ line 2\\ printead{e1}}
% \affiliation{Some University}

\author{\fnms{Matteo} \snm{Fasiolo}\ead[label=e1]{matteo.fasiolo@gmail.com}},
\author{\fnms{Natalya} \snm{Pya}\ead[label=e2]{N.Y.Pya@bath.ac.uk}}
\and
\author{\fnms{Simon N.} \snm{Wood}\ead[label=e3]{s.wood@bath.ac.uk}}
\address{\printead{e1}}
\affiliation{University of Bath}
%\and
%\author{\fnms{???} \snm{???}\ead[label=e2]{???}}
%\address{\printead{e2}}
%\affiliation{???}

\runauthor{Fasiolo, Pya and Wood}

\begin{abstract}
Highly non-linear, chaotic or near chaotic, dynamic models are important in fields such as ecology and epidemiology: for example, pest species and diseases often display highly non-linear dynamics. However, such models are problematic from the point of view of statistical inference. The defining feature of chaotic and near chaotic systems is extreme sensitivity to small changes in system states and parameters, and this can interfere with inference. There are two main classes of methods for circumventing these difficulties: information reduction approaches, such as  
Approximate Bayesian Computation or Synthetic Likelihood and state space methods, such as Particle Markov chain Monte Carlo, Iterated Filtering or Parameter Cascading. 
The purpose of this article is to compare the methods, in order to reach conclusions about how to approach inference with such models in practice. We show that neither class of methods is universally superior to the other. We show that state space methods can suffer multimodality problems in settings with low process noise or model mis-specification, leading to bias toward stable dynamics and high process noise. Information reduction methods avoid this problem but, under the correct model and with sufficient process noise, state space methods lead to substantially sharper inference than information reduction methods. More practically, there are also differences in the tuning requirements of different methods. Our overall conclusion is that model development and checking should probably be performed using an information reduction method with low tuning requirements, while for final inference it is likely to be better to switch to a state space method, checking results against the information reduction approach. 
\end{abstract}


\end{frontmatter}

\section{Introduction} \label{sec: intro}

Non-linear or near-chaotic dynamical systems represent a challenging setting for statistical inference. The chaotic nature of such systems implies that small variations in model parameters can lead to very different observed dynamics. This characteristic alone is enough to invalidate many conventional statistical methods, but in most cases additional complications are present. Firstly, the process under study is generally observed with errors. In addition, many models include a further layer of uncertainty, which we call process stochasticity. In ecology this is often environmental noise, driving the system dynamics. Process stochasticity increases the complexity of the model in a non-trivial way: apart from being unobservable, its presence makes every realized trajectory of the system essentially unique. This is particularly true for chaotic models where any amount of process noise will cause rapid divergence of two paths generated using identical parameters and initial conditions, in sharp contrast to the situation in which dynamics lie on a stable attractor. 

Developing statistical methods that can deal effectively with highly non-linear systems is not simply a matter of theoretical interest, since examples of non-linear or near-chaotic behaviour in ecological systems abound: lemmings \citep{kausrud2008linking}, voles \citep{turchin2000}, mosquitos \citep{yang2008importance}, moths \citep{kendall2005population} and fish \citep{anderson2008fishing}. Similar degrees of non-linearity have been observed in experimental settings, for example: blowflies \citep{nichol57} and flour beetles \citep{desharnais2001chaos}. 

The focus of epidemiologists often differs from that of ecologists. Both groups are concerned with explaining the persistence of the species under study, but epidemiologists and ecologists are often aiming respectively at causing and avoiding its extinction \citep{earn1998persistence}. Despite this divergence in objectives, the mathematical structures used to study population dynamics are often very similar. Hence the role of non-linearities in the population dynamics of infectious diseases has attracted much attention in epidemiology as well. In the context of measles, \cite{grenfell1992chance} and \cite{grenfell1995spatial} describe how the interaction between seasonal forcing and observed heterogeneities, such as age structure or spatial coupling, can result in chaotic or stable dynamics, while \cite{grenfell2002dynamics} address the issue of predictability under a time-series Susceptible Infected Recovered model. More recently \cite{king2008inapparent}, \cite{lavine2013} and \cite{bhadra2011malaria} use non-linear stochastic models with multiple compartments to analyse cholera, pertussis and malaria epidemics, respectively. 

The relation between chaos, statistics and probability theory has been discussed by \cite{berliner1992statistics} and \cite{chan2001chaos}, among others. We have a quite different focus, which is to review and compare the main statistical methods for highly non-linear dynamic models in ecology and epidemiology, investigating the difficulties involved in their use, and attempting to establish the best approach to take in practical applications.

The paper is organized as follows: in Section \ref{sec: MultiModProblem} we show that the likelihood function of simple dynamic models can be intractable in certain areas of the parameter space, while in Section \ref{sec: models} we briefly review the set of statistical methods most useful in the context of non-linear dynamic systems. How these methods deal with the issue discussed in Section \ref{sec: MultiModProblem} is the subject of Section \ref{subsec: multi}. In Section \ref{sec: numComp} we compare the relative performance of these methodologies on a sequence of increasingly realistic (and hence complex)  ecological and epidemiological models. We conclude with a discussion.

\section{Chaos and the likelihood function} \label{sec: MultiModProblem}

To provide a simple example illustrating how the dynamics of an ecological model can challenge conventional statistical approaches, let us consider the noisily observed Ricker map
\begin{equation} \label{eq:obsRicker}
y_t \sim \text{Pois}(\phi n_t),
\end{equation}
\begin{equation} \label{eq:ricker}
n_{t+1} = rn_{t}e^{-n_t+z_{t+1}}, \;\;\;\; z_t \sim N(0,\sigma^2),
\end{equation}
which can be used to describe the evolution in time $t$ of a population $n_t$. Parameter $r$ is the intrinsic growth rate of the population, controlling the dynamics of the system; $\phi$ is a scale parameter. The process noise $z_t$ can be interpreted as environmental noise. 

\begin{figure}
\centering
\includegraphics[scale = 0.25]{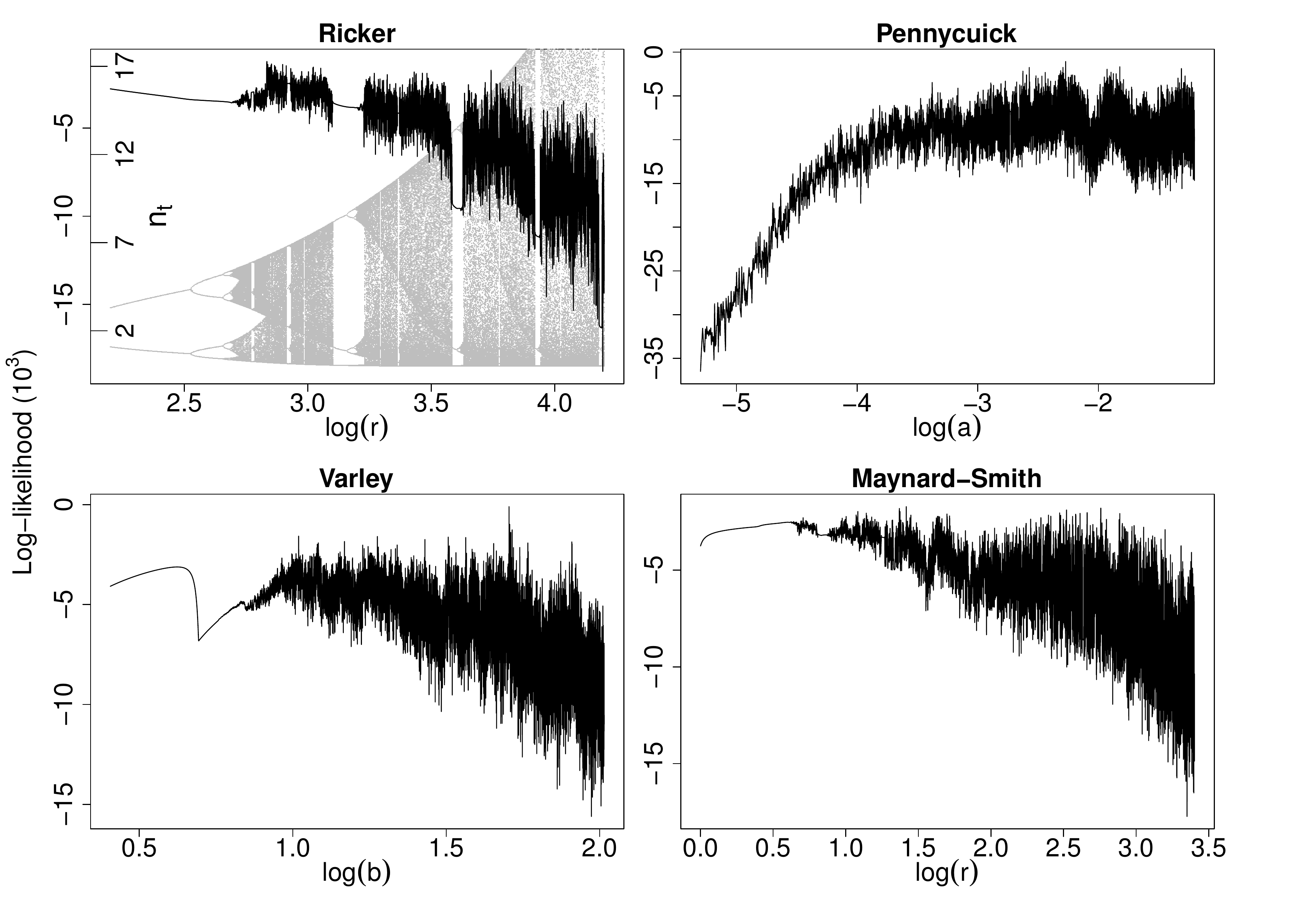}
\caption{Slices of the log-likelihoods of four simple models w.r.t different parameters (black). In each case $\sigma = 0$, hence the likelihoods are analytically available.  For the Ricker map a bifurcation diagram is included (gray).}
\label{fig: rickerSlice}
\end{figure}

Denote with $\bm y_{1:T} = \{\bm y_1, \bm y_2, \dots, \bm y_T\}$ and $\bm n_{1:T} = \{\bm n_1, \bm n_2, \dots, \bm n_T\}$ the observations and hidden state sequence up to time $T$, where $\bm y_t \in \mathbb{R}^{d_y}$ and $\bm n_t \in \mathbb{R}^{d_n} $ for $t \in \{1, \dots, T\}$. Equations (\ref{eq:obsRicker}) and (\ref{eq:ricker}) define a simple state space model (SSM), for which parameter inference is non-trivial: defining $\bm \theta = \{r, \phi, \sigma \}^T$, the likelihood $p(\bm y_{1:T}|\bm \theta)$ is intractable in certain areas of the parameter space. For example, when $\sigma = 0$, the likelihood is analytically available, but extremely irregular for high values of $r$. The plot on the top left of Figure \ref{fig: rickerSlice} shows a transect of the log-likelihood w.r.t. $\log(r)$, obtained using 50 observations, $y_t$, simulated using parameters $\log(r) = 3.8$, $\sigma = 0$ and $\phi = 10$. Given the ragged shape of the log-likelihood, estimating the parameters by maximum likelihood would be very challenging computationally, while having only limited theoretical motivation. Similarly, any standard MCMC algorithm targeting the parameter posterior distributions would hardly mix at all. This behaviour is generic to highly non-linear dynamic systems: Figure \ref{fig: rickerSlice} shows likelihood transects for three more dynamic models, defined in Table \ref{tab:simpleModels}, any of which could be used to make the same points made using the Ricker map, below. 

\def\arraystretch{2}

\begin{table}[t]
\begin{center}
    \begin{tabular}{| l | c| c}
    \hline
    Model Name & Process Equation\\ \hline
    Generalized Ricker & $n_{t+1} = rn_{t}e^{-n_{t}^{\theta} + z_t}$  \\
                         \hline
    Pennycuick & $n_{t+1} = \frac{rn_{t}}{1 + e^{-a(1-n_{t})}}e^{z_t}$  \\ \hline
%    Hassell & $n_{t+1} = \frac{rn_{t}}{(1 + n_{t})^b}e^{z_t}$ &
%              $r = 55$, $b = 100$, $\sigma = 0.3$, $\phi = 1000$ \\ \hline
    Maynard-Smith & $n_{t+1} = \frac{rn_{t}}{(1 + n_{t}^b)}e^{z_t}$  \\ \hline
    Varley & $n_{t+1} = \left\{ \begin{array}{ll}
         rn_{t}e^{z_t} & \text{if $n_{t} \leq c$};\\
         rn_{t}^{1-b}e^{z_t} & \text{if $n_{t} > c$}.\end{array} \right.$   \\
    \hline
    \end{tabular}
\end{center}
\caption{Five simple maps that can show chaotic dynamics. In each case $y_t \sim \text{Pois}( \phi n_t)$ and $z_t \sim N(0, \sigma^2)$.}
\label{tab:simpleModels}
\end{table}
Figure \ref{fig: rickerSlice} reflects the extreme sensitivity of the likelihood of chaotic models to minuscule changes in parameters or process noise. The bifurcation diagram of the Ricker map (grey) shows the possible long term values $n_t$ of the map, as a function of $\log(r)$. While the trajectories oscillate between two values for $\log(r) \approx 2$, increasing $\log(r)$ above $2.5$ leads to a sequence of closely spaced bifurcations, each doubling the periodicity of the map. This period-doubling cascade has a direct effect on the likelihood. Notice that this function is smooth again for values of $\log(r)$ where stable periodic oscillations are recovered. Further increasing $\log(r)$ leads to more period-doubling phases and eventually to chaos. 

\begin{figure}
\centering
\includegraphics[scale = 0.17]{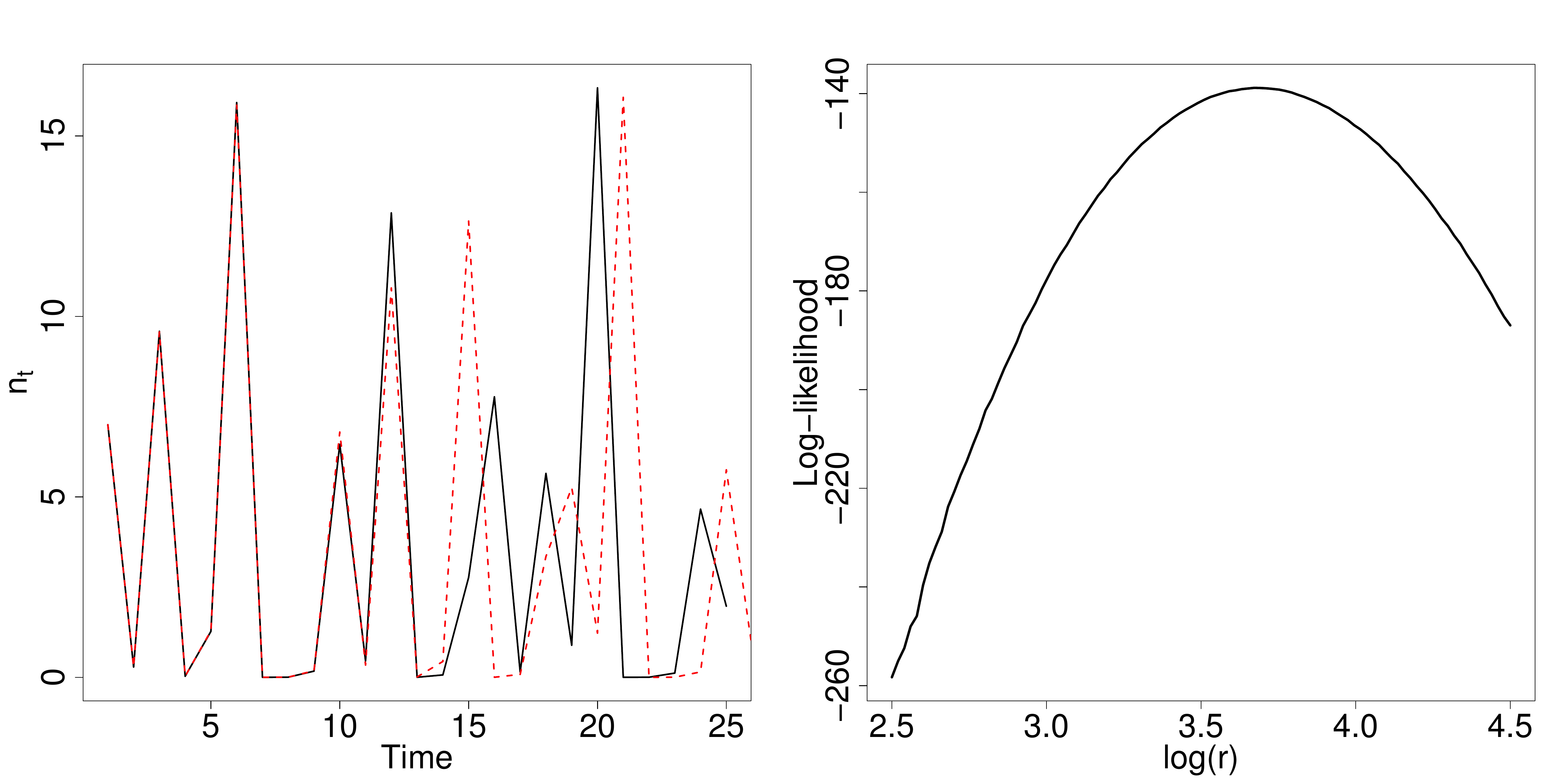}
\caption{Left: two trajectories $\bm n_{1:T}$ of the hidden state, generated using the same initialization, but slightly different values of $\log(r)$. Right: transect w.r.t. $\log (r)$ of the log-likelihood of the Ricker map with $\sigma = 0.3$, estimated using the SIR particle filter. The irregularities at $\log(r) \approx 2.6$ are due to Monte Carlo noise.}
\label{fig: diverPaths}
\end{figure}

Figure \ref{fig: diverPaths} illustrates the origin of this extreme multimodality. We generated two state paths, $\bm n_{1:50}$, using $\sigma = 0$ and the same initial value $n_1 = 7$, but different values of $\log(r)$: 3.8 (black) and 3.799 (red). The two paths are close to each other for the first steps, but the mismatch between them increases with time, and by $t = 15$ the peaks and troughs of the paths do not coincide any more. This sort of divergence of neighbouring trajectories is the defining feature of chaotic dynamics (measured formally in terms of Lyapunov exponents).

The choice $\sigma = 0$ is quite peculiar. What does the likelihood look like when the process dynamics are stochastic? 
\begin{minipage}{0.95\linewidth}
\begin{framed} \label{SIRS}
\begin{center}
\emph{Box 1 \\ Sequential Importance Re-Sampling (SIR) for likelihood estimation}
\end{center} 
This algorithm, originally proposed by \cite{gordon1993novel}, exploits the Markov property to approximate integral (\ref{eq:integralStates}) in $T$ sequential steps. Let $\bm n_0^{1:M}$ be a sample of particles from the prior distribution $p (\bm n_0)$. Then $p(\bm y_{1:T}|\bm \theta)$ is estimated as follows. \\
\vspace{2pt}
For $t = 1$ to T:
\begin{enumerate}[topsep=0pt]
\item For $i = 1,\dots,M$: \\ 
propagate the $i$-th particle forward 
$$
\bm n_t^i \sim p(\bm n_t^i|\bm n_{t-1}^i, \bm \theta),
$$ 
and weight it using the $t$-th observation
$$
w^i = p(\bm y_t|\bm n_t^i, \bm \theta).
$$
\item Estimate the $t$-th likelihood component
$$
\hat{p}(\bm y_t|\bm y_{1:t-1}, \bm \theta) = \frac{1}{M} \sum_{i = 1}^M w^i.
$$
\item Re-sample $\bm n_t^{1:M}$ with replacement, using probabilities proportional to $ \bm w^{1:M}$.
\end{enumerate} 
\vspace{2pt}
Finally, estimate the likelihood by using
\begin{equation*}
\hat{p}(\bm y_{1:T}|\bm \theta) = \hat{p}(\bm y_1|\bm \theta) \prod_{t = 2} ^ T \hat{p}(\bm y_t|\bm y_{1:t-1}, \bm \theta).
\end{equation*}
\end{framed}
\end{minipage}

\vspace{\dimexpr.8pt+\fboxsep\relax}

In this case the likelihood, $p(\bm y_{1:T}|{\bm \theta})$, must be evaluated  by integration
\begin{equation} \label{eq:integralStates}
\begin{split}
p({\bm y}_{1:T}|\bm \theta) & = \int p({\bm y}_{1:T}, {\bm z_{1:T}}|\bm \theta) \, d{\bm z_{1:T}} \\ & = \int p({\bm y}_{1:T}, {\bm n_{1:T}}|\bm \theta) \, d{\bm n_{1:T}}.
\end{split}
\end{equation}
where the second integral is generally the more computationally tractable version.
The plot on the right of Figure \ref{fig: diverPaths} shows a transect of the estimated log-likelihood of the Ricker map w.r.t. parameter $\log(r)$, obtained using the Sequential Importance Re-sampling (SIR) particle filter with $5 \times 10^5$ particles. Box 1 details the main steps of this algorithm, while we refer to \cite{doucet2009tutorial} for a more detailed introduction to particle filters. The observed path ${\bm y}_{1:50}$ has been simulated using $\log(r) = 3.8$, $\sigma = 0.3$ and $\phi = 10$. In sharp contrast with the deterministic case (Figure \ref{fig: rickerSlice}), it appears that the injection of process noise ($\sigma > 0$) into the system has made the likelihood smooth and unimodal. At this point several questions arise: is the likelihood really smooth, as Figure \ref{fig: diverPaths} suggests, or is it possible that the particle filter is hiding the extreme multimodality of Figure \ref{fig: rickerSlice}, so that what we observe in Figure \ref{fig: diverPaths} is an artefact of Monte Carlo integration? If the likelihood is indeed smooth, how did the transition from Figure \ref{fig: rickerSlice} to Figure \ref{fig: diverPaths} occur? How much noise $\sigma$ should be present in order to obtain a smooth likelihood?

\begin{figure}
\centering
\includegraphics[scale = 0.25]{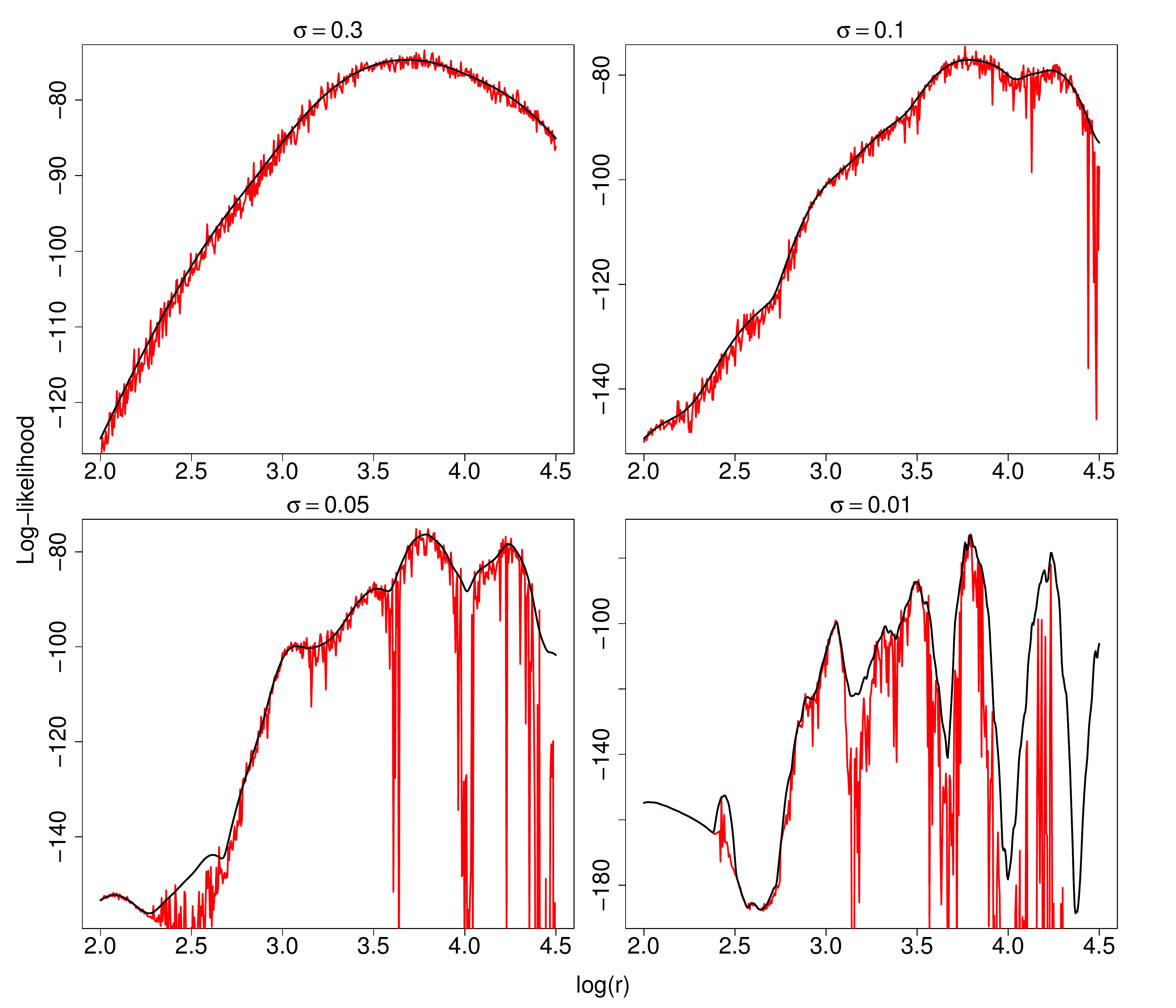}
\caption{Transects of the true log-likelihood (black) of the discrete Ricker map w.r.t. $\log (r)$ for decreasing values of $\sigma$. The red lines are SIR's estimates, using 1000 particles.}
\label{fig: trueLik}
\end{figure}

Checking the reliability of the estimates provided by a particle filter is difficult because, for non-linear and/or non-Gaussian models, Monte Carlo or numerical integration are the only ways to get an approximation to \ref{eq:integralStates}.
To obtain a benchmark against which to compare the estimates of the likelihood provided by the filter, we have therefore discretized the state space of the Ricker map in 500 intervals. In this way we can calculate the likelihood exactly, since the integrations are replaced by efficiently computable summations over all the possible values of the states, as detailed in the Supplementary Material. Obviously, we do not propose discretization as a viable alternative to particle filters, but we want to use a discretized SSM to compare the performance of a particle filter with the true likelihood.  It is interesting to check whether the injection of any amount of noise is sufficient to smooth the likelihood, or whether there is a slow transition from the intractable likelihood shown in Figure \ref{fig: rickerSlice} to the unimodal case of Figure \ref{fig: diverPaths}. Perhaps unsurprisingly, Figure \ref{fig: trueLik} shows that the latter is the case, since as we reduce the process noise the likelihood becomes firstly multimodal and then (for any practical purpose) non-differentiable for very low $\sigma$. Notice that the SIR estimate of the likelihood deteriorates as multi-modality sets in: we will investigate this more fully in Section \ref{subsec: multi}.  

This suggests that there is an area of the parameter space, corresponding to high $\log(r)$ and low $\sigma$, where the likelihood is essentially intractable. For practical purposes it is therefore important to compare the robustness of alternative statistical methods across the parameter space, and to understand how alternative methods behave in the face of this difficulty. In particular, we need to avoid the possibility of concluding that a system's dynamics are relatively stable and noisy, not because they really are, but because that is the only case in which the likelihood is numerically tractable. 

\section{Available statistical methods} \label{sec: models}

The literature contains two main classes of statistical methods for non-linear dynamical systems:
\begin{enumerate}

\item Information reduction: methods that discard the information in the data that is most sensitive to extreme divergence of trajectories, so that fitting objectives become more regular. Two methodologies belonging to this group will be described in Section \ref{sec:infoRed}.

\item State space: these work on the hidden states (${\bf n}_{1:t}$ in Section \ref{sec: MultiModProblem} notation) in order to estimate model parameters and/or the hidden states themselves. Some of these approaches work without modifying the model or the data in any way, by using advanced computational techniques based on particle filtering. We describe two members of this family in Section \ref{sec:stateEstim}.

\end{enumerate}

Given that the main purpose of this work is to consider the applicability and relative performance of these methods in the context of near-chaotic dynamic systems, we will skip over the technical detail whenever they are not essential for the discussion. Obviously our analysis is by no means exhaustive, as we do not examine all the approaches 
%(such as Simulated Quasi-Maximum Likelihood (SQML) \citep{smith1993estimating}) 
that could be applied in this context. In Section \ref{sec:alternatives} we briefly describe some of the alternatives to the methods included in this work. 

\subsection{Approaches based on information reduction} \label{sec:infoRed}

Since the trajectories of near chaotic systems are extremely sensitive to perturbations of parameters or system state, statistical methods that rely on recovering the true system state face a difficult task. At the same time it is often the case that the true state itself is only a nuisance for parameter estimation, and discarding some information regarding the particular observed trajectory might ease the inferential process.

To make this point clearer consider again the Ricker paths in Figure \ref{fig: diverPaths}. Even though the two trajectories, which we indicate with $\bm y_{1:T}$ and $\bm x_{1:T}$, are very different in terms of Euclidean distance $||\bm y_{1:T} - \bm x_{1:T}||$, it is clear that they share some common features. A way around the impossibility of replicating the observed path, even when the simulations use the true or ``best-fitting'' parameters and initial value, is focusing on the relationship between some characteristic features of the data and the unknown parameters. One way of doing this is to transform the observed and simulated data into a set of summary statistics and to base subsequent inferences on these.

In the following we denote by ${\bm y}_{1:T}^0$ the observed path, and with ${\bm s^0} = S({\bm y}_{1:T}^0)$ the vector of observed summary statistic. Often methods based on summary statistics involve two main approximations of the likelihood function. The first is implied by the use of $p({\bm s^0}| {\bm \theta})$ as a proxy for $p({\bm y}_{1:T}^0| {\bm \theta})$, where ${\bm \theta}$ are the model parameters. The second approximation arises from the fact that $p({\bm s^0}| {\bm \theta})$ itself is generally not available analytically and hence it has to be approximated or estimated by simulation.

We will focus on two approaches based on information reduction: Approximate Bayesian Computation (ABC) \citep{beaumont2002approximate, fearnhead2012constructing} and Synthetic Likelihood (SL) \citep{wood2010}. These methods will be outlined in Section \ref{sec:ABC} and \ref{sec:SL}, respectively.

\subsubsection{Approximate Bayesian Computation} \label{sec:ABC}

The main purpose of ABC algorithms is approximating the posterior density $p(\bm \theta|\bm y_{1:T}^0) \propto p(\bm y_{1:T}^0|\bm \theta)p(\bm \theta)$, where $p({\bm \theta})$ is the prior distribution of the model parameters, when the likelihood $p(\bm y_{1:T}^0|\bm \theta)$ is unavailable or intractable. Given that the data is often transformed into a vector of summary statistics, these methods are generally aiming at sampling from $p(\bm \theta|\bm s^0)$ rather than $p(\bm \theta|\bm y_{1:T}^0)$.

An elementary ABC algorithm iterates the following rejection procedure \citep{toni2009approximate}:
\begin{enumerate}
\item Sample a vector of parameters $\bm \theta^i$ from $p(\bm \theta)$.
\item Simulate a path $\bm y_{1:T}^i$ from the model $p(\bm y_{1:T}|\bm \theta^i)$.
\item Transform $\bm y_{1:T}^i$ to a vector of summary statistics ${\bm s^i} = S(\bm y_{1:T}^i)$.
\item Compare ${\bm s^i}$ to the observed statistics ${\bm s^0}$ using a pre-specified distance measure $d(\cdot,\cdot)$. If $d({\bm s^i}, {\bm s^0}) \leq \epsilon$, where $\epsilon \geq 0$, accept $\bm \theta^*$ otherwise reject it.
\end{enumerate}
The output of this algorithm will be distributed according to
\begin{equation*}
p(\bm \theta)p \{ d({\bm s},{\bm s^0} ) < \epsilon |\bm \theta \} \propto p \big \{\bm \theta | d({\bm s},{\bm s^0} ) <\epsilon \big \},
\end{equation*}
which approximates the posterior density, $p(\bm \theta |\bm s_0)$, for sufficiently small
 $\epsilon$. In practice simple rejection ABC is replaced with MCMC or Sequential Monte Carlo (SMC) algorithms.  

\subsubsection{Synthetic Likelihood} \label{sec:SL}

Similarly to ABC, this method can be used for problems where the likelihood is intractable, but it is still possible to simulate from the model. The main difference between ABC and SL is how $p({\bm s^0}|{\bm \theta})$ is approximated. ABC does not rely on any distributional assumption on ${\bm s}$, while SL assumes that, approximately,
\begin{equation*}
S(\bm y) \sim N({\bm{\mu}_{\theta}},  \bf \Sigma_{\theta}).
\end{equation*}
Briefly, a pointwise estimate of the synthetic likelihood at $\bm \theta$ can be obtained as follows:
\begin{enumerate}
\item Simulate $N$ datasets ${\bm y}_{1:T}^1, \dots, {\bm y}_{1:T}^N$ from the model $p({\bm y}_{1:T}|\bm \theta)$.
\item Transform each dataset ${\bm y}_{1:T}^i$ into a $d$-dimensional vector of summary statistics $S({\bm y}_{1:T}^i)$.
\item Calculate the sample mean $\hat{\bm \mu}_{\bm \theta}$ and covariance matrix $\hat{\bm \Sigma}_{\bm \theta}$ of the statistics (often robustly).
\item Estimate the synthetic likelihood
\begin{equation*}
\begin{split}
\hat{p}({\bm s}^0|{\bm \theta}) & =  (2\pi)^{-\frac{d}{2}}|\hat{\bm \Sigma}_{\bm \theta}|^{-\frac{1}{2}}  \\ 
&  \times \exp \bigg \{-\frac{1}{2}({\bm s}^0 - \hat {\bm \mu}_{\bm \theta})^T \hat {\bm \Sigma}_{\bm \theta}^{-1} ({\bm {s}}^0 - \hat {\bm \mu}_{\bm \theta}) \bigg \}.
\end{split}
\end{equation*}
\end{enumerate}
Hence SL explicitly provides point estimates of $p({\bm s^0}|{\bm \theta})$. This estimator can be used within Markov chain Monte Carlo (MCMC) algorithms approximately targeting $p({\bm \theta}|{\bm s^0})$, or within an optimizer aiming at maximizing the synthetic likelihood.

\subsection{State space methods} \label{sec:stateEstim}

If discarding information through the use of summary statistics is not desirable, then it is necessary to deal with the hidden states explicitly. As previously stated, calculating the likelihood of SSMs involves integrating the hidden states $\bm n_{1:T}$ out of the joint density $p({\bm y}^0_{1:T}, {\bm n}_{1:T}|{\bm \theta})$. The SIR particle filter can be used to obtain a Monte Carlo estimate of the likelihood, by employing a sequential integration scheme. The use of a sequential approach allows filters to direct the simulated trajectories of the hidden states toward values that are consistent with the observations. This feature is particularly attractive in the context of near-chaotic models, where simulated paths diverge rapidly (recall Figure \ref{fig: diverPaths}). In this work we mainly focus on algorithms based on the SIR scheme, but  many other approaches are available. For example, it is possible to use algorithms that sample directly from the joint posterior density of parameters and hidden states, thus circumventing the estimation of the likelihood. For detailed overviews see \cite{andrieu2010particle} and \cite{doucet2000sequential}.

Here we consider three state space approaches, two of which are based on particle filtering. In Section \ref{sec: pmcmc} we describe a sampler belonging to the family of Particle Markov chain Monte Carlo (PMCMC) methods \citep{andrieu2010particle}, while in Section \ref{sec: IF} we introduce the Iterated Filtering (IF) algorithm \citep{ionides2011iterated}. We consider the Parameter Cascading approach proposed by \cite{ramsay2007parameter} in Section \ref{sec:ramsay}.

%Figure (\ref{fig: sliceNoise}) illustrates not only why doing the integration is problematic, but it also suggests a possible way forward. In fact, we see how the degree of multimodality decreases as we move from time 1 to time T. This is because if we change $e_1$ all the subsequent steps of the path will be modified, while if we change $e_ {47}$ only the remaining four steps of the paths will be perturbed ($T = 50$). Obviously changing  $e_ {50}$ modifies only that point of the path, and hence the transect of the joint probability w.r.t. $e_ {50}$ (not shown) is perfectly unimodal. Therefore intuitively one could suppose that there should be a way to work out the integral using a step by step procedure.

\subsubsection{Particle Marginal Metropolis-Hastings sampler} \label{sec: pmcmc}

Filters such as the SIR algorithm can provide point estimates $\hat{p}({\bm y}^0_{1:T}|{\bm \theta})$ of the likelihood, which ideally converge to the true likelihood as the number of simulations increases. \cite{andrieu2010particle} proposed to use these estimates of the likelihood to set up a Particle Marginal Metropolis-Hastings (PMMH) algorithm, which can be used to sample from the posterior distribution of the parameters. The algorithm is formed by the following steps:
\begin{itemize}
\item Step 1: Initialization $i = 0$. \\Given an estimate or a guess of the parameters $\bm \theta_0$, estimate the likelihood $p({\bm y}^0_{1:T}|\bm \theta_0)$ using a particle filter.
\item Iteration $i \geq 1$:
\begin{enumerate}
\item sample a new vector of parameters ${\bm \theta}^*$ from a transition kernel $K({\bm \theta}^*|{\bm \theta}_{i-1})$.
\item Using a particle filter estimate the likelihood $\hat{p}({\bm y}^0_{1:T}|\bm \theta^*)$.
\item With probability
$$
min \bigg \{ 1,\frac{\hat{p}({\bm y}^0_{1:T}|\bm \theta^*)p(\bm \theta^*)}{\hat{p}({\bm y}^0_{1:T}|\bm \theta_{i-1})p(\bm \theta_{i-1})}\frac{K({\bm \theta}_{i-1}|{\bm \theta}^*)}{K({\bm \theta}^*|{\bm \theta}_{i-1})}  \bigg \},
$$
set ${\bm \theta}_{i} = {\bm \theta}^*$, otherwise set  ${\bm \theta}_{i} = {\bm \theta}_{i-1}$.
\end{enumerate}
\end{itemize}
This algorithm is exact in the sense that, despite the use of noisy estimates of $p(\bm y^0_{1:T}|\bm \theta)$ in the acceptance step, it will generate a dependent sample from $p(\bm \theta|\bm y^0_{1:T})$. The conditions under which this occurs are detailed in \cite{andrieu2009pseudo}.

\subsubsection{Iterated filtering} \label{sec: IF}

The IF algorithm uses particle filters to provide approximate Maximum Likelihood estimates of the unknown parameters. As shown by \cite{ionides2006inference}, by including the unknown parameters in the state space and running a filtering operation, it is possible to estimate the gradient of the likelihood function, which can then be used within an optimization routine. In more detail, \cite{ionides2006inference} treat the parameters as if they were following a multivariate random walk
\begin{equation}
{\bm \theta}_{t} = {\bm \theta}_{t-1} + {\bm \psi}_{t} ~~ \text{with} ~~   {\bm \psi}_{t} \sim N({\bm 0}, \sigma^2 {\bm \Sigma}).
\end{equation}
With this choice we have that
$$
E({\bm \theta}_{t}|{\bm \theta}_{t-1}) = {\bm \theta}_{t-1}, ~~~~~  Var({\bm \theta}_{t}|{\bm \theta}_{t-1}) = \sigma^2 {\bm \Sigma},
$$
$$
E({\bm \theta}_{0}) = \hat{\bm \theta} ~~~~\text{and}~~~~~~  Var({\bm \theta}_{0}) = c^2 \sigma^2 {\bm \Sigma},
$$
where $\sigma$ and $c^2$ are two variance multipliers, $\hat{\bm \theta}$ is an initial estimate, while ${\bm \Sigma}$ is typically a diagonal matrix, giving the respective scale of the parameters. 

The main result underlying the IF algorithm is
\begin{equation} \label{eq:basicIF}
\lim_{\sigma^2 \to 0} \sum_{t = 1}^T \bm V_t^{-1}( \hat{\bm \theta}_t -  \hat{\bm \theta}_{t-1}) = \nabla \log{p({\bm y}^0_{1:T}|{\bm \theta})},
\end{equation}
where
$$
\hat{{\bm \theta}}_t = E({\bm \theta}_t|{\bm y}^0_{1:t}) ~~ \text{and} ~~ {\bm V}_t = Var({\bm \theta}_t|{\bm y}^0_{1:t}),
$$
can be estimated using the SIR particle filter. The IF algorithm is composed of the following steps:
\begin{itemize}
\item Choose initial value $\hat{\bm \theta}^{(0)}_0$, parameters $\sigma^2$, $c^2$, $\bm \Sigma$, $\alpha \in (0, 1)$ and number of iterations $M$.
\item Iterate for j in $1,\dots,M$:
\begin{enumerate}
\item Set $\sigma_j = \alpha^{j-1}$. Estimate $\hat{{\bm \theta}}_t^{(j)}$ and ${\bm V}_t^{(j)}$, for $t = 1, \dots, T$, using a particle filter.
\item Update the parameter estimate 
$$
\hat{{\bm \theta}}^{(j+1)}_0 = \hat{{\bm \theta}}^{(j)}_0 + {\bm V}_1^{(j)}\sum_{t = 1}^T ({\bm V}_t^{(j)})^{-1}( \hat{{\bm \theta}}_t^{(j)} -  \hat{{\bm \theta}}_{t-1}^{(j)}).
$$
\end{enumerate}
\item Then $\hat{{\bm \theta}}^{(M+1)}_0$ is an approximate Maximum Likelihood estimate of the parameters.
\end{itemize}
Notice that, as long as $\sigma > 0$, IF will not be fitting the original model, which will be recovered as $\sigma \rightarrow 0$. \cite{ionides2011iterated} give results concerning the theoretical foundation of IF and describe how slowly $\sigma$ has to decrease to assure convergence. 

\subsubsection{Parameter Cascading} \label{sec:ramsay}

In the context of Ordinary Differential Equations (ODEs), \cite{ramsay2007parameter} proposed an approach to parameter estimation which can be adapted to the discrete-time models, such as the Ricker map. The estimation procedure is a nested optimization problem with three levels. Given $\lambda$ and a current estimate  $\hat{\bm \theta}$, the hidden states are estimated by minimizing an inner criterion
\[\arraycolsep=1.4pt\def\arraystretch{1.8}
\begin{array}{lll}
\bm n_{1:T}^{\hat{\bm \theta}} & = & \underset{\bm n_{1:T}}{\text{argmin}}\,
J(\bm n_{1:t}|\hat{\bm \theta}, \lambda)  \\
& = & \underset{\bm n_{1:T}}{\text{argmin}} \bigg \{ - \sum_{t=1}^T \log p(\bm y^0_t|\bm n_t, \hat{\bm \theta}) + \lambda \psi (\bm n_{1:T}|\hat{\bm \theta}) \bigg \}, 
\end{array}
\]
where 
$$
\psi (\bm n_{1:T}|\hat{\bm \theta}) = \sum_{t = 1}^T \big \{\bm n_t - E(\bm n_t|\bm n_{t-1}, \hat{\bm \theta}) \big \}^2,
$$
quantifies deviations of the estimated state from the model, while $\lambda$ determines the trade-off between data fitting and model compliance. The parameters are estimated using the higher level criterion
\[\arraycolsep=1.4pt\def\arraystretch{1.8} 
\begin{array}{lll} 
\hat{\bm \theta} & = & \underset{\bm \theta}{\text{argmin}}\, H(\bm \theta|
\bm n_{1:T}^{\hat{\bm \theta}}, \lambda) \\ 
& = & \underset{\bm \theta}{\text{argmin}}\, \big \{ - \sum_{t=1}^T \log p(\bm y^0_t|\bm n_t^{\hat{\bm \theta}}, \bm \theta) \big \}.
\end{array}
\]
A further level can be added in which an outer grid search is used to select $\lambda$. This method is especially useful for exploring multimodality problems in Section \ref{subsec: multi}.

%We do not include this method in the comparison presented in section \ref{sec: numComp}, in part because of the reasons illustrated in section \ref{subsec: multi}.

%\section{Information reduction versus state space methods} \label{sec: IRvsSM}

\subsection{Alternative approaches} \label{sec:alternatives}

The methods described in the preceding sections represent a subset of those that could be used in the context of parameter estimation for non-linear state space models. Here we discuss some of the alternatives, describe their relation with the methods described above and detail our reasons for not including them in this work.

There exist a large variety of particle-filtering-based methods that can be used to obtain approximate Maximum Likelihood (ML) estimates of the static parameters, such as \cite{andrieu2005online}, \cite{andrieu2003online}, \cite{malik2011particle}, \cite{poyiadjis2011particle} and \cite{nemeth2013particle}. IF belongs to this class of methods, and we chose to include it, rather than some of the alternatives, in this work because (i) it is theoretically justified, as detailed in \cite{ionides2011iterated}, (ii) it is has been tested on a variety of complex models, such as those described in \cite{king2008inapparent}, \cite{he2010plug} and \cite{bhadra2011malaria}, which are of direct interest to applied researchers in ecology and epidemiology, and (iii) the computational cost of a score function estimate is $O(M)$ in the number of particles, which, to our best knowledge, is the state of the art. Hence we argue that, by including IF, this work should adequately cover this class of methods.

Notably, this work does not include MCMC methods for parameter identification, such as those proposed by \cite{carlin1992monte}, \cite{geweke2001bayesian}, \cite{polson2008practical} and \cite{niemi2010adaptive}. One reason for this is that highly non-linear models, such as those considered here, are often characterized by strong dependencies between states and static parameters. Under such circumstances, implementing an efficient MCMC sampler requires the design of adequate conditional proposal densities, which is not trivial for non-linear non-Gaussian models \citep{andrieu2010particle, kantas2014particle}. In addition, the model presented in Section \ref{subsec: Cholera} is a discretized version of a continuous time model, where the discretization error was limited by using a large number of intermediate states between each pair of observations. Sampling this enlarged state space using standard MCMC methods would be challenging, because the convergence rate of such schemes can be arbitrarily slow if the amount of augmentation is large \citep{roberts2001inference}. With the exception of Parameter Cascading, all the methods described in our work are less affected by this problem, because the intermediate states are simply simulated forward using $p(\bm n_t|\bm n_{t-1},\bm \theta)$. This ``plug-and-play'' property is one of the reasons behind popularity of these methods \citep{ionides2011iterated}.

Apart from PMCMC and MCMC algorithms, the methods proposed by \cite{Kitagawa1998} and \cite{liu2001combined} could also be used to sample the posterior distribution of $\bm \theta$. Analogously to IF, these filters include the parameters in the state space, and perturb them using an artificial noise process. Even though \cite{liu2001combined} counteract the resulting over-dispersion of the posterior by shrinking the perturbed parameters toward their mean, this does not entirely eliminate the information loss, if the posterior is far from Gaussian. Hence, in this work we preferred to target $p(\bm \theta|\bm y_{1:T})$ using PMMH, because of the convergence guarantees detailed in \cite{andrieu2009pseudo}. However, the computational cost of PMMH is fairly high, and the filter of \cite{liu2001combined} might be able to sample a close approximation to $p(\bm \theta|\bm y_{1:T})$, using far fewer filtering operations.

Finally, the versions of IF and PMMH used here are based on the SIR algorithm, as described in \cite{gordon1993novel} and \cite{doucet2000sequential}. More sophisticated filters, such as those proposed by \cite{pitt1999filtering} and \cite{klaas2012toward}, might provide more accurate estimates of the likelihood, or of $\nabla p(\bm y_{1:T}|\bm \theta)$ in the context of IF. Similarly, it might be possible to improve upon the MCMC implementation of ABC and SL used in Section \ref{sec: numComp}, by using more sophisticated SMC samplers \citep{toni2009approximate} or Gaussian Processes \citep{meeds2014gps}, respectively. We do not explore these possibilities here, because doing so would increase the complexity of this work, without adding much to its main results.

\section{Multimodality and state space methods} \label{subsec: multi}
%(XXX: The code for all this is in ``Hidden.modes.discrete.R'')

If the presence of process noise smooths the likelihood sufficiently, then methods that discard information should be outperformed by those that retain it. However, we can not generally prove that the likelihood for any particular model is smoothed and, as shown in Section \ref{sec: MultiModProblem}, there exist models for which smoothing is only partial, and may be inadequate, when process noise is low. In this section we further investigate the impact of multimodality on state space methods, and show that information reduction methods can reduce the associated problems.     

\begin{figure}
\centering
\includegraphics[scale = 0.2]{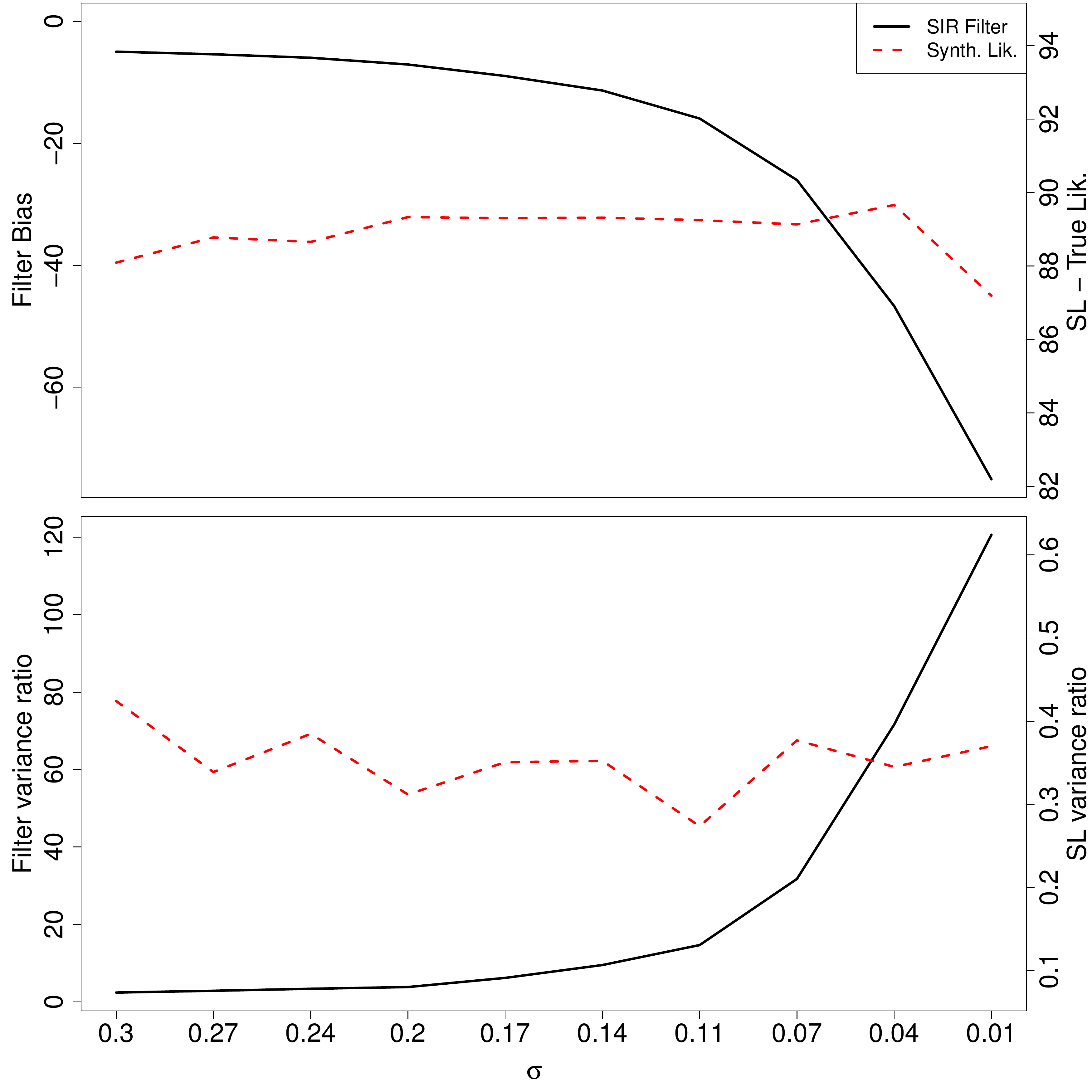}
\caption{Top: average difference between the full likelihood and the estimated full (solid) or synthetic likelihood (dashed) as a function of $\sigma$, obtained using respectively the SIR filter and SL. Bottom: ratio between the sample variance of estimated full (black line) or synthetic (broken red line) likelihoods and the true likelihood for several values of $\sigma$.}
\label{fig: decreasNoise}
\end{figure}

In order to evaluate the accuracy of the likelihood estimates given by the SIR algorithm for different levels of noise, we used the discretized SSM described in Section \ref{sec: MultiModProblem} and in the Supplementary Material. We chose ten levels of process noise in the interval $\sigma \in [0.01, 0.3]$. For each level we  simulated $1000$ paths using the Ricker map, with $\log(r) = 3.8$, $\phi = 0.5$, and  evaluated the likelihood of each of them at the true parameters. Figure \ref{fig: decreasNoise} shows the results.

The plot on the top shows that, as the process noise decreases, the average bias of the likelihood estimated by the filter (solid) increases in absolute value. Indeed, while the true log-likelihood (not shown) is roughly constant ($\approx -70$) for different levels of $\sigma$, the mean filter's estimates drop from -65 for $\sigma = 0.3$ to -140 for $\sigma = 0.01$. The strong dependence between likelihood bias and $\sigma$ suggests that a sampler using these likelihood estimates will never explore areas of the parameter space where $\sigma$ is low.  In addition, any model comparison criterion based on the biased likelihood estimates is unreliable.

On the bottom of Figure \ref{fig: decreasNoise} we plotted the ratios between sample variance of the likelihood estimated by the filter and the sample variance of the true likelihood for each value of $\sigma$, that is
$$
\frac{\hat{\text{Var}} \big \{ \log \hat{p}({\bm y}_{1:50}|{\bm \theta}) \big \} }
{\hat{\text{Var}}\big \{\log p({\bm y}_{1:50}|{\bm \theta})\big \}}.
$$
From the plot we see that the variance of the estimated log-likelihood increases exponentially as $\sigma$ decreases, suggesting that Monte Carlo variability of the integration procedure dwarfs sampling variation for low $\sigma$. This has implications for algorithms based on particle filters: with such noisy likelihood estimates the PMMH algorithm will have an extremely low acceptance rate \citep{doucet2012efficient}, while the IF procedure will become quite unstable, due to the high variability of the estimated gradients.

The broken lines in Figure \ref{fig: decreasNoise}, show corresponding quantities for the synthetic likelihood, obtained using the set of 13 summary statistics proposed by \cite{wood2010} and reported in the Supplementary Material.
%\begin{itemize}
%\item the autocovariances of the path $y_1,\dots,y_T$ up to lag 5;
%\item the mean population $\bar{y}$;
%\item the number of zeros observed;
%\item the coefficients of the regression $y_{t+1}^{0.3}=\beta_1y_t^{0.3}+\beta_2y_t^{0.6}+Z_t$;
%\item the coefficients of a cubic regression of the ordered differences $y_t-y_{t-1}$ on their observed values.
%\end{itemize} 
Interestingly, both the average and the variance of the synthetic likelihood estimates remain roughly constant for different degrees of process noise. This suggests that the SL approach is quite robust to the level of process noise in the system, as it gives stable estimates also when the process dynamics are near-deterministic. On the other hand, the variance of the synthetic likelihood is lower than that of the true likelihood for any $\sigma$, which might be a consequence of the information loss.

Note that to use synthetic likelihood when the system is (close to) deterministic, the initial values of the simulated paths have to be randomized ($N_1 \sim \text{Unif(0.1, 5)}$), otherwise the variances of the summary statistics can be close to zero for very low process noise. Random initial values are consistent with the information reduction philosophy: inference should be robust to the particular values of the hidden states. In this context we are confident that ABC, being based on summary statistics, would perform similarly to SL.

\begin{figure}
\centering
\includegraphics[scale = 0.25]{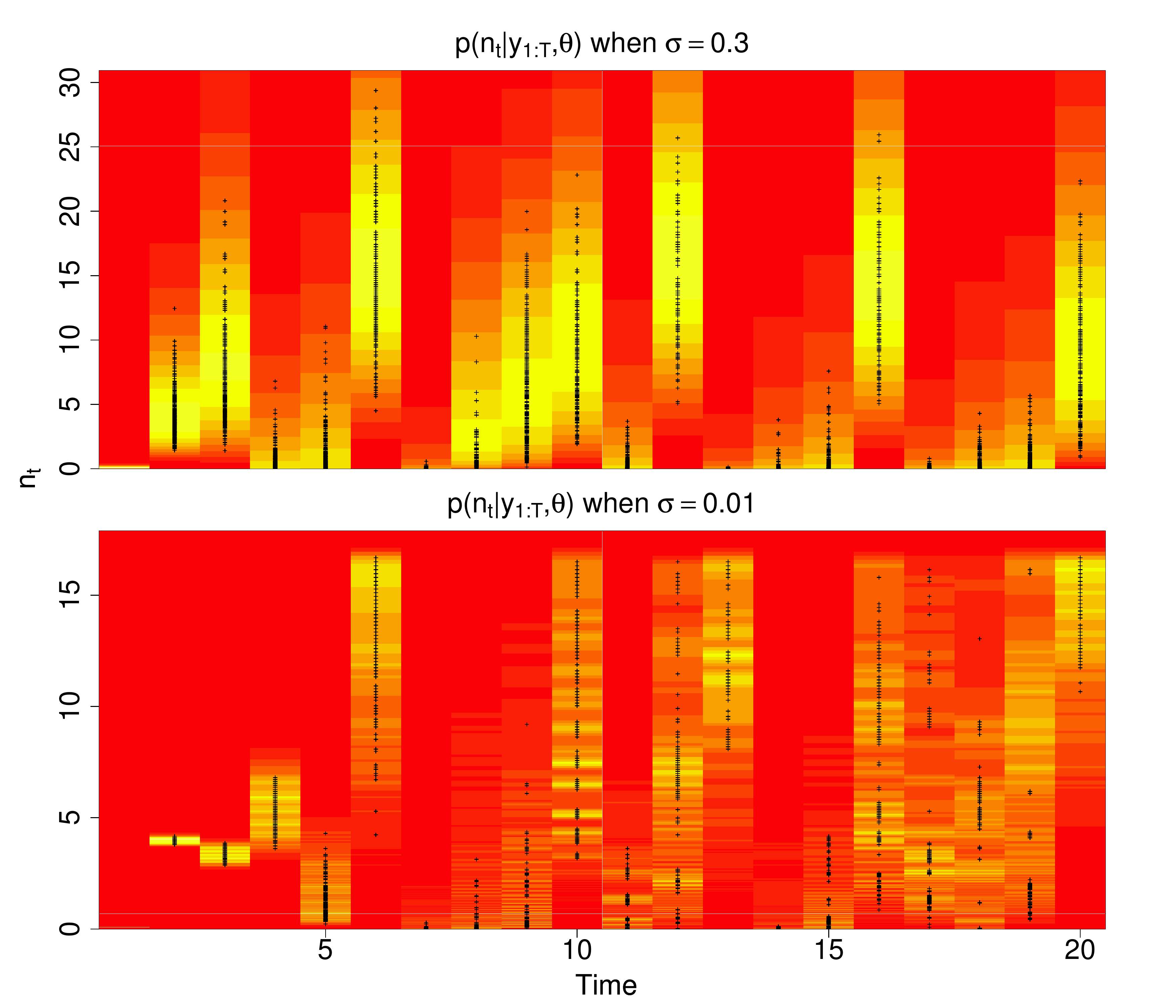}
\caption{Filtering densities $p(n_t | \bm y_{1:t}, \bm \theta)$ for a single Ricker path generated using $log(r) =3.8$, $\phi = 10$ and $\sigma = 0.3$ (top) or $\sigma = 0.01$ (bottom).}
\label{fig: Low_vs_high_noise}
\end{figure}

Figure \ref{fig: Low_vs_high_noise} shows why the SIR algorithm is struggling to estimate the log-likelihood when $\sigma$ is very low. Each of the 20 columns in the top image represents the true filtering density $p(n_t | {\bm y}_{1:t}, {\bm \theta})$ at each time step, when $\sigma = 0.3$. Areas of high density are represented in yellow, while area of lower density are coloured in red. With this level of process noise the filtering densities are smooth and unimodal, so the filter places the particles around each mode, thus providing a reliable estimate of the likelihood. 
%(the estimation of the likelihood relies on the approximation of the filtering densities, see Box 1). 
In contrast, the image on the bottom of Figure \ref{fig: Low_vs_high_noise} shows that for very low process noise the filtering densities are unimodal in the first couple of time step, but then they break into narrow multiple modes. Because of the irregularity of the filtering densities, the quality of the particle approximation is poor in this case (see time 19 in particular). The filter struggles to explore all the important modes of the filtering distributions, and hence the resulting estimates of the log-likelihood are very variable.

So Figure \ref{fig: Low_vs_high_noise} helps to explain the variability in performance of the particle filter approach seen in Figures \ref{fig: trueLik} and \ref{fig: decreasNoise} as the process noise level changes. For models capable of showing chaotic or near-chaotic dynamics, there will be areas of the parameter space where the likelihood is highly multimodal. In these areas particle filtering methods will struggle to estimate the likelihood. In such situations most of the likelihood-based asymptotic theory will not be applicable, and even if it was possible to sample the corresponding parameter posterior exactly, it would not be obvious how the results should be interpreted. Hence we argue that in such situations the use of approaches based on information reduction, which can provide a smooth proxy to likelihood, might be preferable from both a methodological and practical point of view.

\begin{figure}
         \begin{center}
             \includegraphics[draft=false,height=9.2cm,width=7cm,clip=false,angle=270]
                             {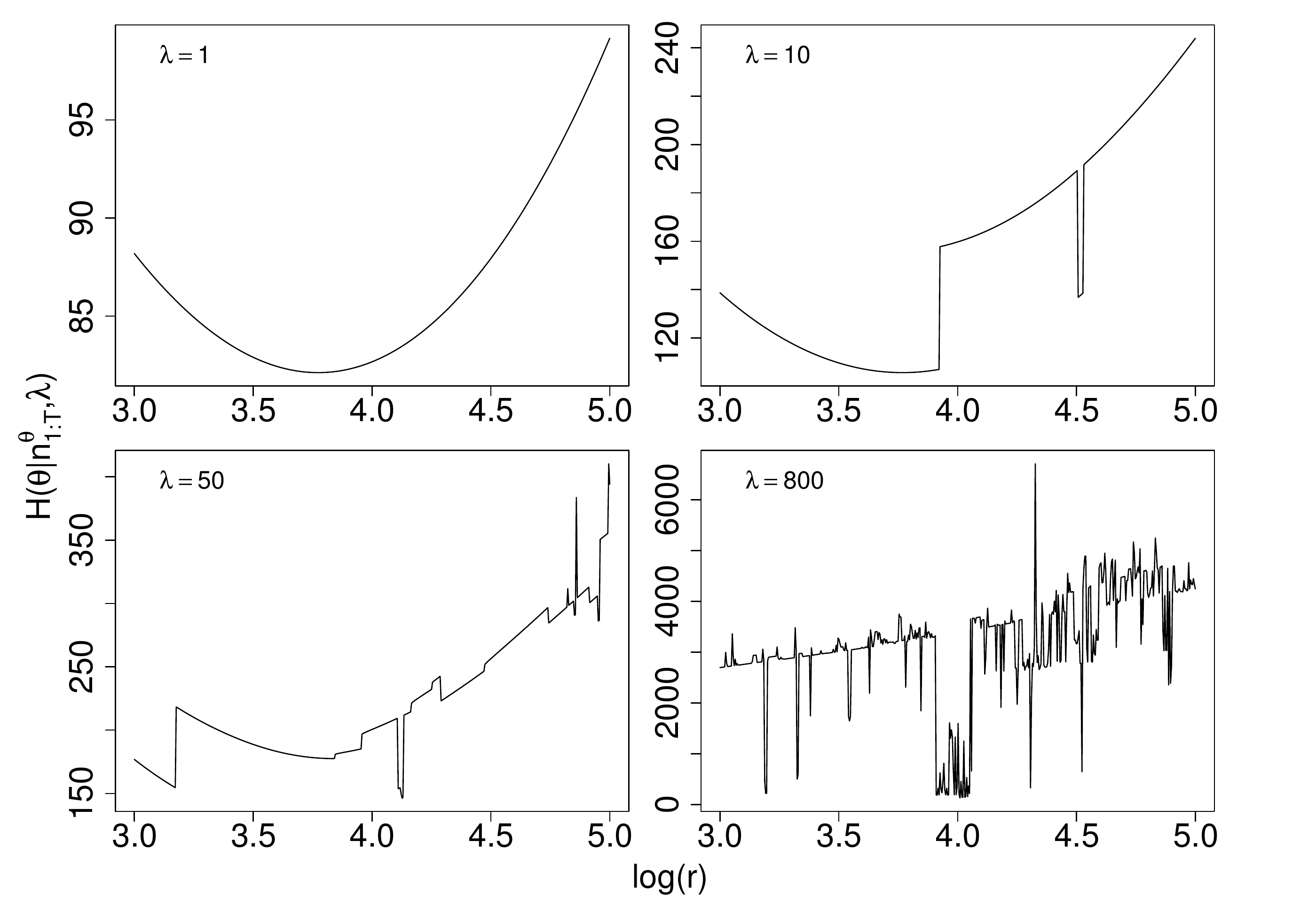}
          \end{center}
        \caption{Transects of $H(\bm \theta|\bm n_{1:T}, \lambda)$ w.r.t. $\log(r)$, as $\lambda$ increases.}
        \label{transect_H}
\end{figure}

To emphasise that the issue of multimodality is generic to the state space approach, rather than being specific to filtering, or a particular filtering implementation, or our discretized state space example, we illustrate how Parameter Cascading can encounter similar problems on the unmodified Ricker model. Figure \ref{transect_H} shows transects of the parameter fitting objective function, $H(\bm \theta| \bm n_{1:T}^{\bm \theta}, \lambda)$, (see Section \ref{sec:ramsay}) with respect to $\log(r)$ for four values of $\lambda$, and show that this function becomes more irregular as $\lambda$ increases. For large $\lambda$, which is appropriate when $\sigma$ is low, this hinders the optimization and makes estimating $\bm \theta$ problematic. In the following we illustrate that jumps in the objective function correspond to transitions between modes of the objective function for the state, $J(\bm n_{1:T}|\bm \theta, \lambda)$.

\begin{figure}
         \begin{center}
             \includegraphics[draft=false,height=9.2cm,width=8cm,clip=false,angle=270]
                             {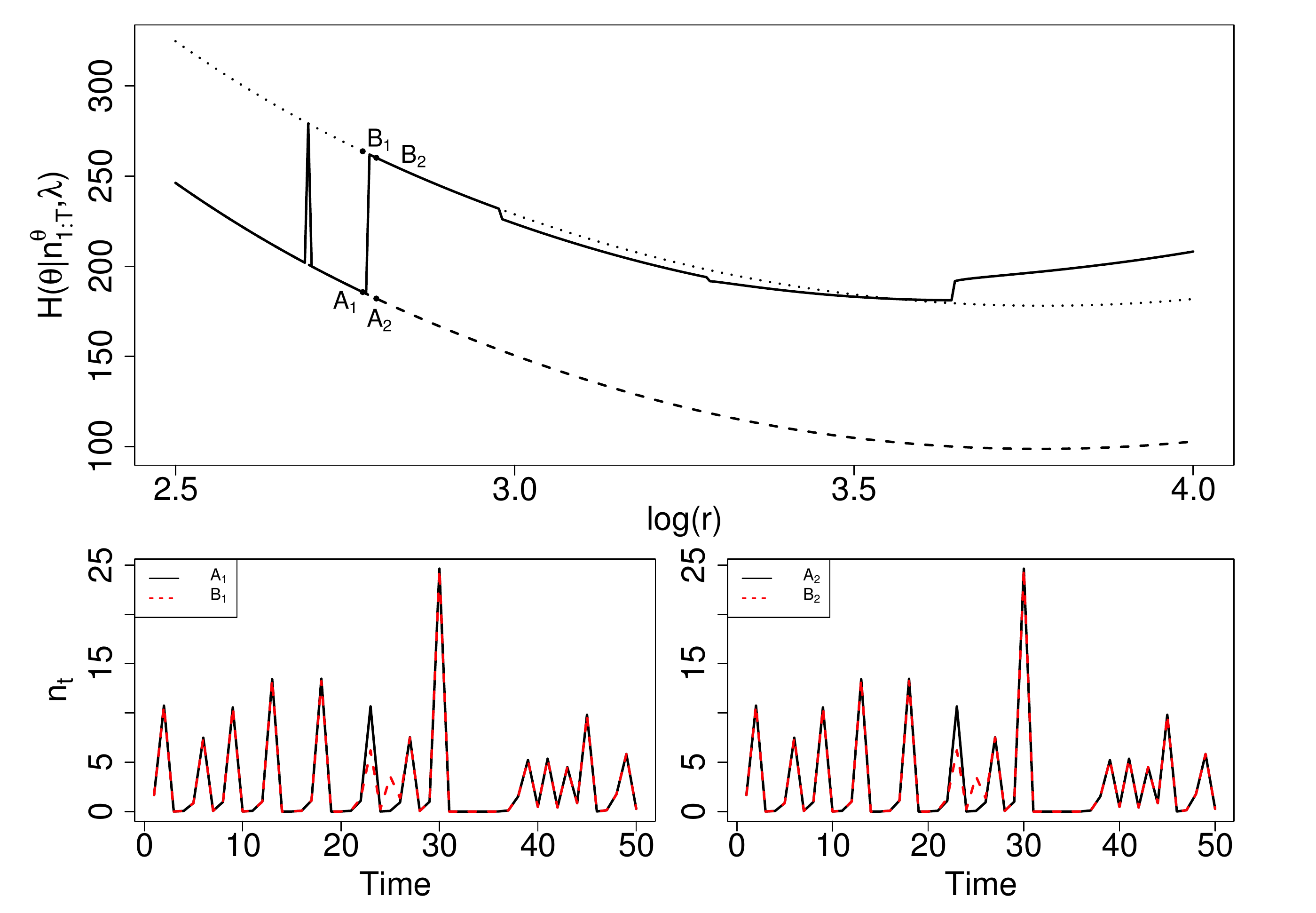}
          \end{center}
        \caption{Top: transects of $H(\bm\theta |\lambda, {\bf n}_t)$ with respect to $\log(r)$. Bottom: paths corresponding to two points 1 or 2 along the $\log (r)$ axis and to modes A or B in the state space.}
        \label{multi}
  \end{figure}

The upper plot of Figure \ref{multi} shows other transects of $H(\bm \theta|\bm n_{1:T}^{\bm \theta}, \lambda)$, for $\lambda = 65$. The solid line was obtained using the same initial value $\bm n_{1:T}^{\bm \theta} = \bm y_{1:T} / \phi$ for each value of $\log(r)$. The dashed lines show the $H(\bm \theta|\bm n_{1:T}^{\bm \theta}, \lambda)$ curves corresponding to two different modes of $J(\bm n_{1:t}|\bm \theta, \lambda)$ and have been obtained by carefully tracking of the modes. We refer to these modes as A and B. The plots on the bottom of Figure \ref{multi} represent the estimated hidden states $\bm n_{1:T}^{\theta}$ corresponding to two values of $\log(r)$ and to each mode. This shows that the same value of $\log(r)$ leads to two different modes in the state space, depending on the initialization. The similarity between the pairs A1-A2 and B1-B2 shows that these initialization-dependent modes are persistent along $\log(r)$. 

\section{Performance comparison} \label{sec: numComp}

In the last section we saw that state space methods for highly non-linear dynamic models can encounter difficulties in some regions of parameter space. Information reduction approaches might then  be preferable, if they show little practical reduction in inferential performance when the dynamics are less problematic. This section therefore compares the relative performance of the statistical approaches presented by employing them to fit several models, using both simulated and real datasets.

\subsection{Example 1: Simple chaotic maps with sufficient noise} \label{subsec: simpleMaps}

\begin{figure}
\centering
\includegraphics[scale = 0.25]{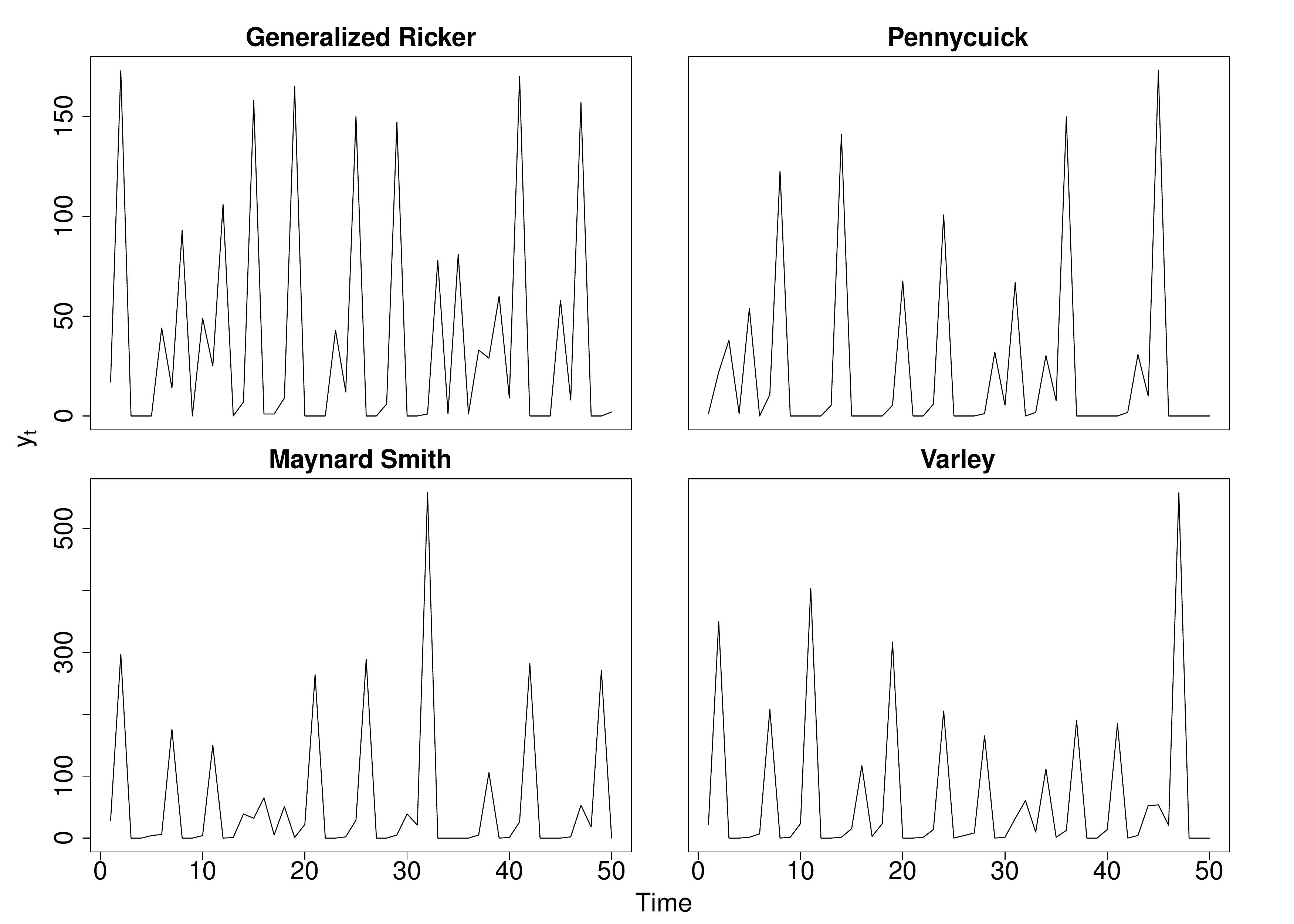}
\caption{Trajectories simulated using the four models described in Table \ref{tab:simpleModels}.}
\label{fig: fourPaths}
\end{figure}

Here we consider the models summarized in Table \ref{tab:simpleModels}, in addition to the Ricker map. The parameter values of each model, reported in the Supplementary Material, have been chosen so that the simulated paths show similar chaotic dynamics (Figure \ref{fig: fourPaths}).

The data consist of 50 simulated paths $\bm y_{1:T}$, where $T = 50$, from each model. All paths were used to estimate the parameters using each method. For SL and for the ABC-MCMC algorithm of \cite{marjoram2003markov} we have used $3 \times 10^4$ iterations to sample the posterior of each path. The PMMH algorithm had an extremely low acceptance rate unless the likelihood of the latest accepted position was re-estimated at each MCMC step. This doubled the computational effort, and hence we used only $1.5 \times 10^4$ iterations for this method. To check if recomputing the likelihood was biasing the results in favour of PMMH, we have implemented a version of SL (labelled SL-R) that uses the same approach. For SL and ABC we have discarded $5000$ iterations as burn-in, while for PMMH and SL-R $2500$ iterations were discarded. For IF we have used $3000$ optimization steps.

At each MCMC step, SL and PMMH estimated the (synthetic) likelihood by using $500$ simulations from the model, while IF used $5000$ simulations at each step of optimization step. ABC simulates only one sample at each step, but we stored an iteration every $500$. Notice that, with this set-up, SL, SL-R, PMMH and ABC used the same number of simulations ($1.5 \times 10^7$) from the model in order to fit each of the 250 simulated datasets. Given that the methods have very different implementation, basing the comparison on the number of simulations from the model, rather than CPU time, ensures fairness.  

We used proper uniform priors for all parameters. IF does not support the use of priors, so we interpreted the priors as box constraints for the optimization. All methods were initialized at the same starting values which, together with the priors and other details, are included in the Supplementary Material.

To choose the tolerance and the distance measure used by ABC-MCMC, we employed the following approach. For each model, we simulated $L = 10^5$ parameter vectors, $\bm \theta_1, \dots, \bm \theta_L$, from $p(\bm \theta)$ and the corresponding statistics vectors, $\bm s_1, \dots, \bm s_L$, from $p(\bm s|\bm \theta)$. As distance measure $d(\bm s, \bm s^0)$ we used $(\bm s - \bm s^0)^T \bm Q^{-1} (\bm s - \bm s^0)$, where $\bm Q = diag(\hat{\bm \Sigma})$, with $\hat{\bm \Sigma}$ being the empirical covariance matrix of the simulated statistics.  We then calculated the distances $d(\bm s_i, \bm s^0)$, for $i = 1, \dots, L$, and we chose $\epsilon$ so that only $0.1\%$ of the distances fell below this threshold.

We evaluated the accuracy of different approaches in term of squared errors between point estimates and the true parameters.  While IF provided point estimates directly, ABC, SL and PMMH give dependent samples from the (approximate) parameter posteriors. Hence for the latter group of methods we have used the posterior means as point estimates. 

\begin{figure}
\centering
\includegraphics[scale = 0.32]{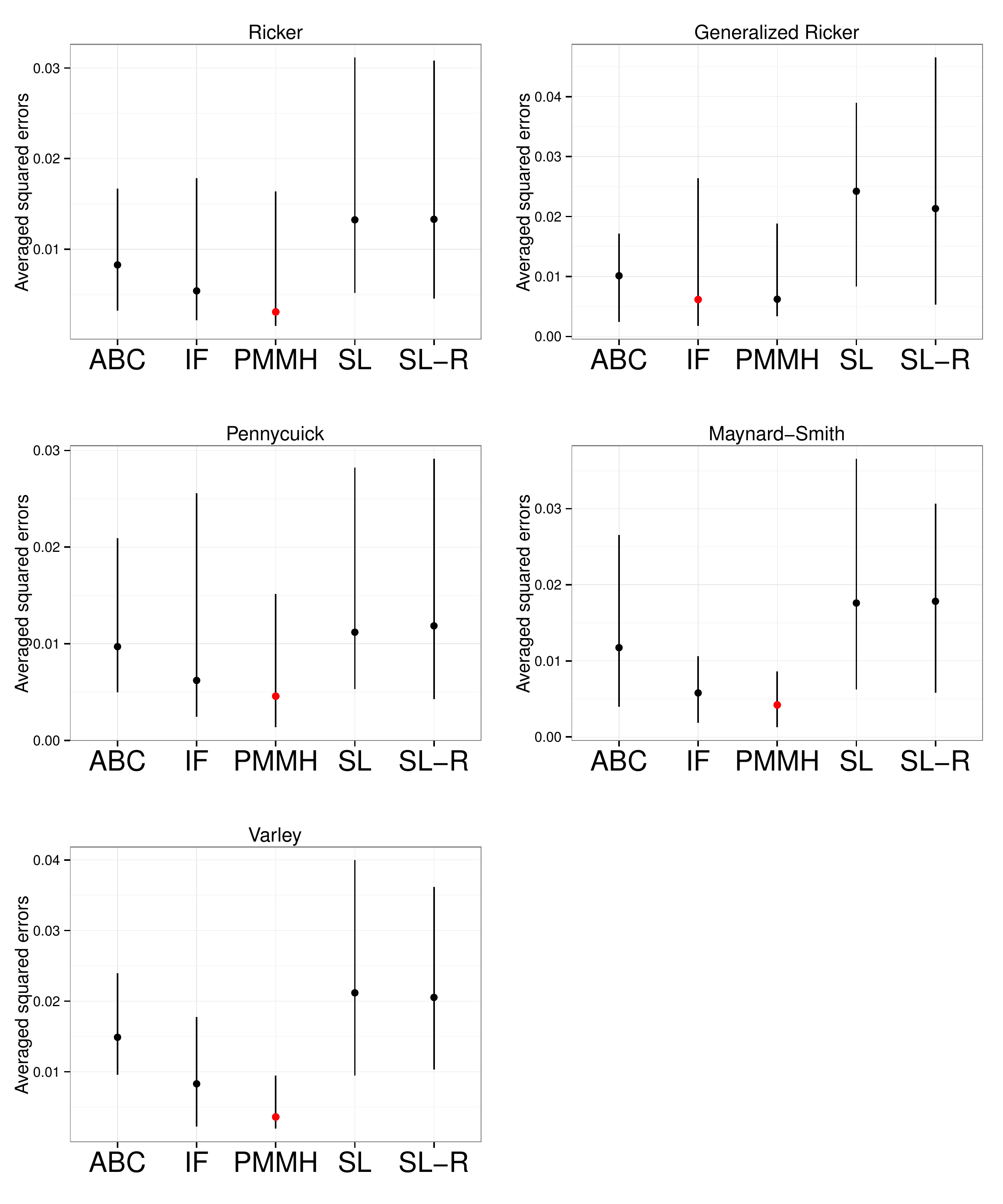}
\caption{Medians and Inter-Quartile Ranges of the averaged squared errors for each model and method.}
\label{fig: compPlot}
\end{figure}

The Supplementary Material reports the median squared errors for each model-method-parameter combination. Here we have summarized the results in Figure \ref{fig: compPlot} which represents, for each model and method, the median and Inter-Quartile Range of the squared errors, averaged geometrically across the parameters. Let $m$, $k$, $j$ and $i$ be the indexes of model, method, dataset and parameter respectively, the average squared errors are then given by
$$
\bar{e}_{j}^{m, k} = \bigg \{ \prod_{i = 1}^{p_m} \big ( \hat{\theta}_{j, i}^{m, k} - \theta_{i}^m \big )^2 \bigg \} ^{\frac{1}{p_m}},
$$
where $p_m$ is the parameter count for model $m$.

Figure \ref{fig: compPlot} shows that, on this set of simple models, methods based on particle filtering consistently outperform methods based on information reduction. The performance of IF and PMMH is quite similar, and the differences in average squared errors between these two methods might be due to the different type of point estimates used. ABC-MCMC seems to perform better that either SL or SL-R for all models. This performance gap might be attributable to the normal approximation used by SL, to the bias entailed by estimating $p(\bm s_0|\bm \theta)$ using a finite sample or simply to particular set-up we have used for the experiment.

Tuning the tolerance and the scaling matrix of ABC-MCMC required little extra effort for the simple models used here. However, the tuning tends to be much more laborious under more complex models, such as described in the following sections. In particular, when the number of unknown parameters is high, training $\epsilon$ and $\bm Q$ using simulations from the prior can be very inefficient, especially if the prior contains little information. Hence, for complex models, tuning $\epsilon$ and $\bm Q$ might require a more sophisticated approach, possibly involving some degree of manual intervention. From this practical perspective, SL is at an advantage, because the summary statistics are scaled automatically using $\bm \hat{\Sigma}_{\bm \theta}$, while no tolerance needs to be chosen.

The clear result here is that, given sufficient noise, the information reduction methods have noticeably worse performance than the state space methods for these simple toy models. In the next sub sections we turn to more realistic examples. In order to limit the computational and programming effort we will restrict our attention to PMMH and SL: that is, one method from each of the two inferential philosophies. We chose SL rather than ABC, because the former method requires much less tuning, as discussed above. We selected PMMH over IF, because PMMH and SL have very similar MCMC implementations, which should limit the influence of other implementational confounders on the results of the comparison.

\subsection{Example 2: Nicholson's blowflies} \label{subsec: blow}

\begin{figure}
\centering
\includegraphics[scale = 0.32]{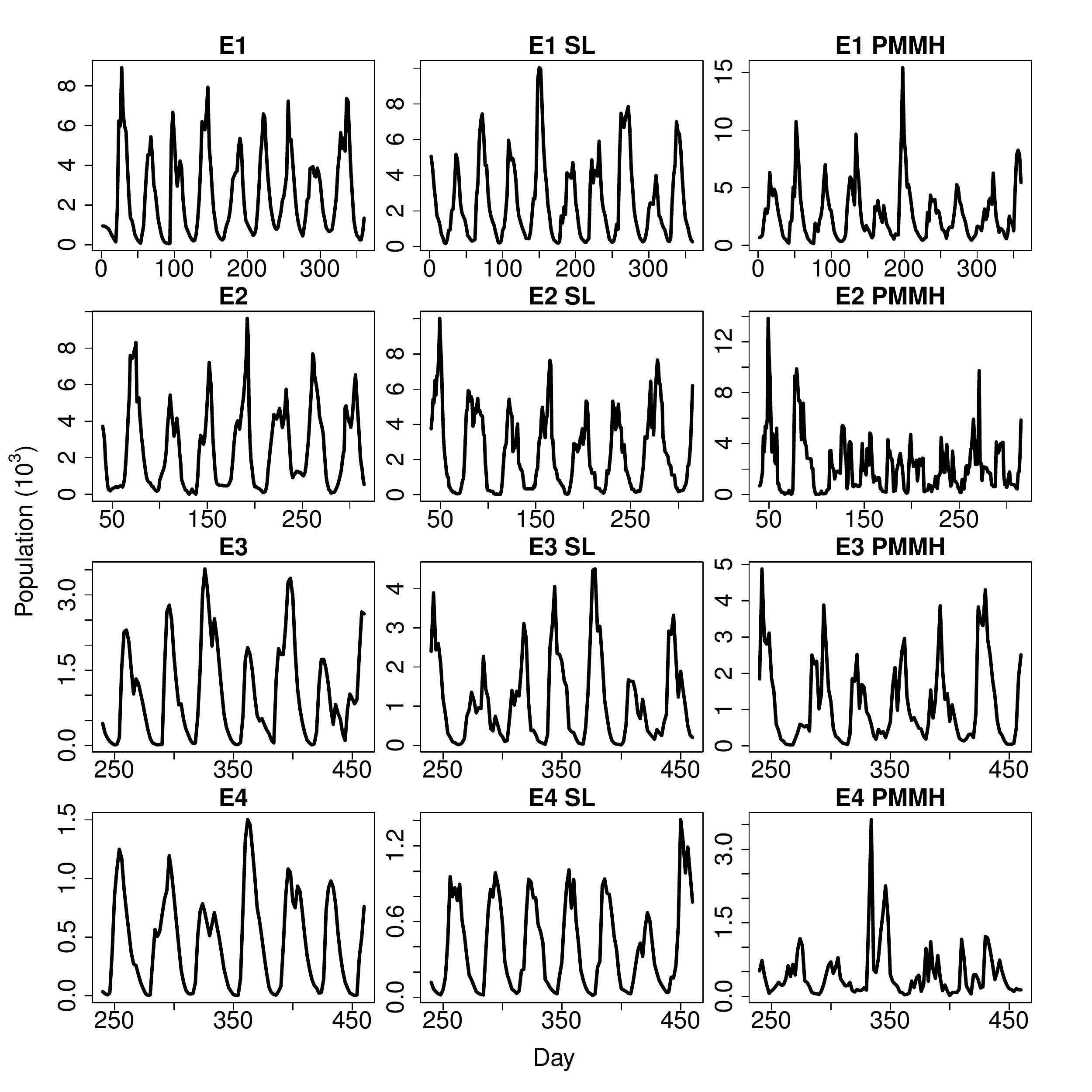}
\caption{Left column: the datasets reported by \cite{nichol54} and \cite{nichol57}. Central and right columns: paths simulated from model \ref{eq: blowDiff} using parameters equal to the posterior means obtained by fitting the four datasets using SL and PMMH. %The trajectories do not have to be in phase as they have different initializations and the initial $2000$ days of simulation were discarded.
}
\label{fig:dataBlow}
\end{figure}

In this section we consider the results, reported by \cite{nichol54} and \cite{nichol57}, of a series of laboratory experiments meant to elucidate the population dynamics of sheep blowfly \emph{Lucilia cuprina} under resource limitation. Blowflies develop in four successive stages: eggs, larvae, pupae and adults. Feeding occurs only in the larval and adult stages. In two of the experiments (E1 and E2) the larvae had unlimited resources, while the adults had unlimited access to sugar and water, but were provided with a limited amount of protein, which is required for egg production. In another two experiments (E3 and E4) the larvae were supplied respectively with a moderately and severely restricted amount of food, while adults had unlimited resources. The resulting population dynamics are shown in the left column of Figure \ref{fig:dataBlow}.

\subsubsection{The model}

A model potentially capable of explaining the observed dynamics of this population was proposed by \cite{gur1980nich}, and it is represented by the following delayed differential equation
\begin{equation} \label{eq: blowDiff}
\frac{dn(t)}{dt} = Pn(t-\tau)e^{-\frac{n(t-\tau)}{n_0}} - \delta n(t),
\end{equation}
where $n$ represents the adult population, while $P$, $\tau$, $n_0$ and $\delta$ are parameters. In order to fit the model to the available datasets \cite{wood2010} proposed a discretized version of equation (\ref{eq: blowDiff}) and added a stochastic component to its deterministic structure. More precisely, he proposed the following model
\begin{equation} \label{eq: blowModel}
n_{t} = r_t + s_t,
\end{equation}
where 
$$
r_t \sim \text{Pois}(Pn_{t-\tau}e^{-\frac{n_{t-\tau}}{n_0}}e_t),
$$
represents delayed recruitment process, while
$$
s_t \sim \text{binom}(e^{-\delta\epsilon_t}, n_{t-1}),
$$
denotes the adult survival process. Finally, $e_t$ and $\epsilon_t$ are independent gamma distributed random variables, with unit means and variances equal to $\sigma_p^2$ and $\sigma_d^2$ respectively.

\subsubsection{Comparison using simulated data}

In order to verify the accuracy of SL and PMMH for the blowfly model, we have tested them on simulated data. Before moving to the results, notice that model (\ref{eq: blowModel}) does not include any measurement noise: the number of blowflies $n_t$ is assumed to be perfectly observed. This means that the model is not a SSM, hence it cannot be fitted using methods based on particle filtering directly. Our solution has been to introduce an artificial measurement process, when fitting the model using PMMH. More precisely, we use the following log-normal observational process
$$
\log{y_t} \sim \text{N}(\log{n_t}, \sigma_o^2),
$$
where the value of $\sigma_o$ was predetermined, not estimated. Notice that, because of this modification, PMMH is fitting the wrong model and this procedure can be seen as an importance sampling ABC procedure, where $\sigma_o$ plays the role of the tolerance. See \cite{dean2011} for more details about the use ABC procedures in the context of SSMs with intractable observational processes. Despite having introduced an artificial measurement process, we have decided to avoid estimating the initial values $n_1, \dotsm, n_{\tau}$ when using PMMH, but we have fixed their values to that of the first $\tau$ observations. 

\begin{table*}[ht]
\begin{center}
\begin{tabular}{rllllll}
  \hline
 & $\delta$ & P & $n_0$ & $\sigma_p^2$ & $\tau$ & $\sigma_d^2$ \\ 
  \hline 
SL0 & 0.00598(0.83) & 0.01686(0.83) & 0.01032(0.79) & 0.05845(1) & 0.00123(0.92) & 0.18568(0.96) \\ 
  PMMH0 & 0.004(0.67) & 0.01176(0.88) & 0.00509(0.88) & 0.30579(0.58) & 0.00042(0.92) & 1.73206(0.17) \\ 
  p-value  & 0.414 & 0.197 & 0.01 & 0.359 & 0.03 & $<0.001$ \\ 
  Best & PMMH0 & PMMH0 & PMMH0 & SL0 & PMMH0 & SL0 \\ 
  \hline
 SL1 & 0.00286(0.83) & 0.01929(0.75) & 0.00836(0.88) & 0.0634(1) & 0.00088(0.96) & 0.18419(1) \\ 
  PMMH1 & 0.00165(0.88) & 0.00416(0.92) & 0.00069(0.92) & 0.03322(1) & 1e-05(1) & 0.02965(0.96) \\ 
  p-value & 0.123 & 0.006 & $<0.001$ & 0.058 & 0.006 & $<0.001$ \\ 
  Best & PMMH1 & PMMH1 & PMMH1 & PMMH1 & PMMH1 & PMMH1 \\ 
   \hline 
\end{tabular}
\end{center}
\caption{MSEs(coverage) of the log-parameters for SL and PMMH for the blowflies model for realistic (0) and optimistic (1) starting values. The p-values for the differences in log-absolute errors have been calculated using t-tests.}
\label{tab:blowMse}
\end{table*}

For the comparison we have simulated 24 datasets of length $T = 200$, using parameter values $\delta$ = 0.16, $P$ = 6.5, $n_0$ = 400, $\sigma_p^2$ = 0.1, $\tau$ = 14, $\sigma_d^2$ = 0.1. We have then estimated the parameters with both methods, using $2\times10^4$ MCMC iteration and $1000$ simulation from the model at each step. The choice of $\sigma_o$ was critical for the performance of PMMH. Obviously we would like $\sigma_o$ to be as small as possible, but lowering it increases the variance of the importance weights and, in turn, of the estimated likelihood. In particular, if PMMH was initialized far from the true parameters, $\sigma_o$ had to be increased in order to avoid particle depletion. Hence, we decided to include the results (PMMH0 and SL0) obtained using a realistic initialization ($\delta = 0.1$, $P = 4$, $n_0 = 200$, $\sigma_p^2 = 0.2$, $\tau = 10$, $\sigma_d^2 = 0.2$) and the results obtained by initializing the chains at the true parameters. In the first case $\sigma_o$ was fixed to $0.05$, while in the second to $0.01$. For all parameters we used flat priors and for SL we used the set of 16 summary statistics proposed by \cite{wood2010} for this model. We report these details in the Supplementary Material.

The running time of the two algorithms was very similar. In particular, when computed on one core of a 3.60GHz i7-3820 CPU, a single estimates of $p(\bm y^0|\bm \theta)$ and $p(\bm s^0|\bm \theta)$ took around 0.25 and 0.29 seconds, respectively.

The resulting Mean Squared Errors (MSEs) of the log-parameters are reported in Table \ref{tab:blowMse}. The table includes the p-values for differences in MSEs, which clearly show that PMMH is more accurate when the lower value of $\sigma_o$ is used. On the other hand, in the more realistic setting the performance of the two procedure is more comparable, as PMMH underestimates both $\sigma_p^2$ and $\sigma_d^2$, while SL performs slightly worse than PMMH on the remaining parameters.

%\begin{figure}
%\centering
%\includegraphics[scale = 0.4]{stabSimul.pdf}
%\caption{Stability plot for the blowfly model, obtained by fitting the 24 simulated datasets using SL (black) and PMMH (red) for the realistic initialization (white circle).}
%\label{fig: stabSimul}
%\end{figure}

\subsubsection{Results using Nicholson's datasets}

\begin{figure}
\centering
\includegraphics[scale = 0.35]{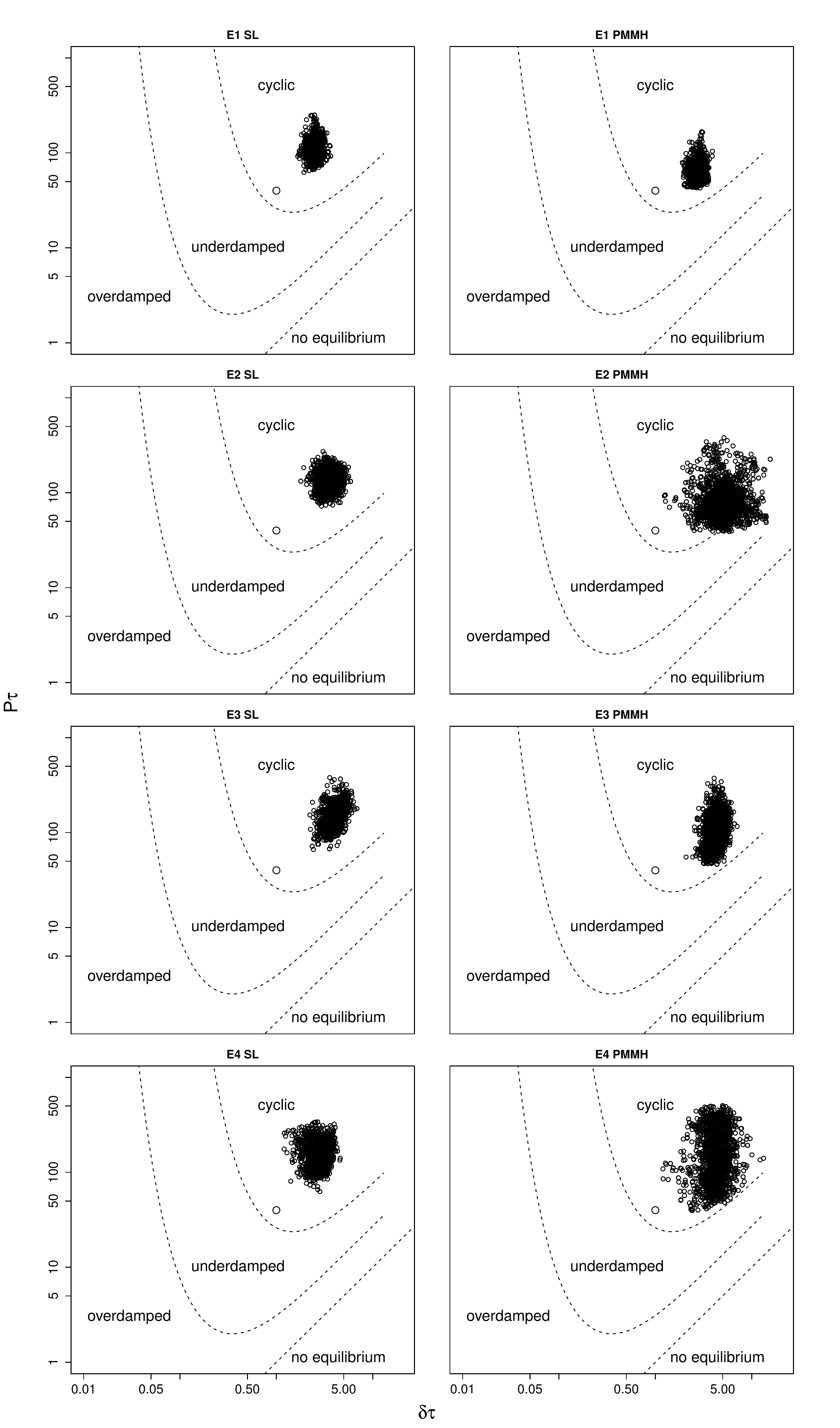}
\caption{Stability plots for the blowfly model, obtained by fitting Nicholson's datasets using SL and PMMH. The black dots are 2000 values of the $P \tau$ and $\delta \tau$ randomly sampled from each MCMC chain. The white circle represents the initial value used for SL.}
\label{fig: stabTrue}
\end{figure}

Fitting Nicholson's datasets was relatively straightforward with SL, and we used the same initial values ($\delta$ = 0.16, $P$ = 6.5, $n_0$ = 400, $\sigma_p^2$ = 0.1, $\tau$ = 14, $\sigma_d^2$ = 0.1) for each dataset. Using this initialization was not possible for PMMH, as we would be forced to use values of $\sigma_o$ as high as $0.2$, in order to avoid failures in the Monte Carlo integration step (i.e. all importance weights were going to zero). Hence we initialized PMMH using values obtained through preliminary runs of SL on the four datasets. Still, we were forced to use values of $\sigma_o$ equal to $0.1$ for the second dataset and $0.05$ for the others. For each dataset we used $3 \times 10^4$ MCMC iterations, of which the first $5000$ were discarded as burn-in. The (synthetic) likelihood was estimated using $1000$ particles or simulated paths at each step.

Figure \ref{fig: stabTrue} shows the stability diagrams for model (\ref{eq: blowModel}), for each combination of dataset and fitting procedure. These plots show how the stability properties of the system depend on the parameter combinations $P \tau$ and $\delta \tau$. All posterior samples obtained through SL lay strictly in the cyclic region of the parameter space, indicating that observed oscillation of blowfly population are due to intrinsic blowfly biology, rather than stochastic perturbation of the system \citep{wood2010}. On the other hand, the posteriors samples given by PMMH, in particular those corresponding to datasets E2 and E4, are closer to the under-damped region, where the oscillations are driven by the stochasticity rather than intrinsic effects. With the exception of E1, the PMMH posteriors are more dispersed, which is attributable to the high estimates of noise parameters $\sigma_d^2$ and $\sigma_p^2$, as shown in Table \ref{tab:blowEstim}.

\begin{table}[ht]
\begin{center}
\begin{tabular}{lllllll}
  \hline
 & $\delta$ & P & $n_0$ & $\sigma_p^2$ & $\tau$ & $\sigma_d^2$ \\ 
  \hline
E1 SL & 0.17 & 7.57 & 395.30 & 0.70 & 14.44 & 0.47 \\ 
  E1 PMMH & 0.19 & 4.45 & 653.93 & 1.54 & 14.82 & 0.30 \\ 
  \hline
  E2 SL & 0.22 & 8.70 & 407.61 & 0.21 & 15.95 & 1.77 \\ 
  E2 PMMH & 0.37 & 6.26 & 576.30 & 2.35 & 15.02 & 3.47 \\ 
  \hline
  E3 SL & 0.29 & 10.48 & 184.38 & 0.64 & 14.62 & 0.55 \\ 
  E3 PMMH & 0.28 & 7.71 & 229.32 & 1.56 & 15.18 & 0.53 \\
  \hline
  E4 SL & 0.22 & 12.81 & 59.16 & 0.71 & 12.91 & 0.55 \\ 
  E4 PMMH & 0.30 & 12.10 & 88.33 & 2.42 & 14.46 & 1.23 \\ 
   \hline
\end{tabular}
\end{center}
\caption{Posterior means for model (\ref{eq: blowModel}), obtained by fitting each of Nicholson's dataset using either SL or PMMH.}
\label{tab:blowEstim}
\end{table}

Figure \ref{fig:dataBlow} compares the observed trajectories with those simulated from the model, using parameter values equal to the posterior means estimated by SL and PMMH. While using parameter values estimated through SL gives trajectories that are qualitatively similar to the observed ones in all cases, using the parameters estimated through PMMH gives a poor match for datasets E2 and E4.

\begin{figure}[t]
\centering
\includegraphics[scale = 0.30]{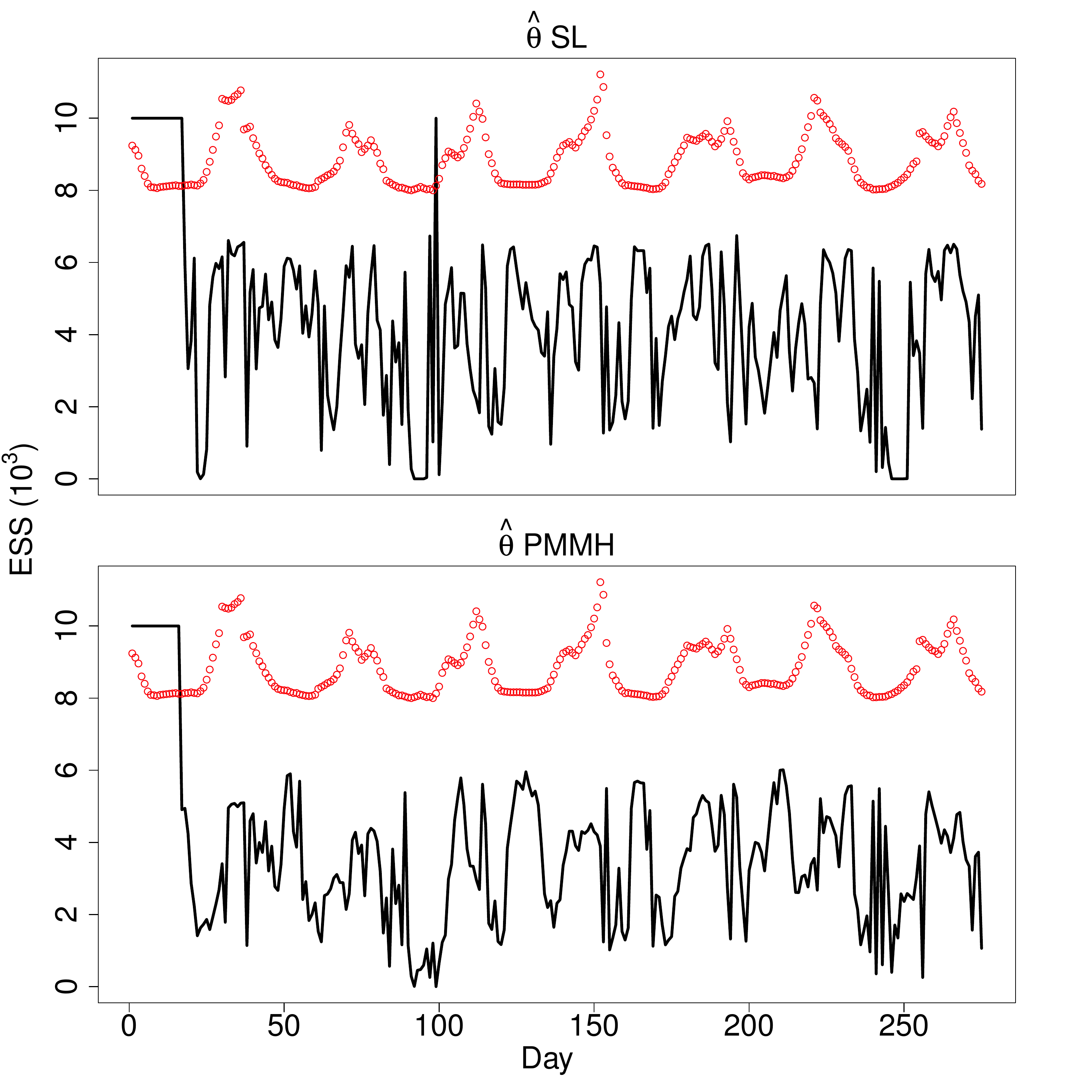}
\caption{Dynamics of the ESS (black line) for the E2 dataset (red points), using parameters equal to the posterior means given by SL (top) and PMMH (bottom). For the first $\tau$ steps the ESS is equal to the number of particles, because we have set $n_i = y_i$, for $i = 1, \dots, \tau$, as stated in the main text.}
\label{fig:ESS}
\end{figure}

\begin{figure}[t]
\centering
\includegraphics[scale = 0.25]{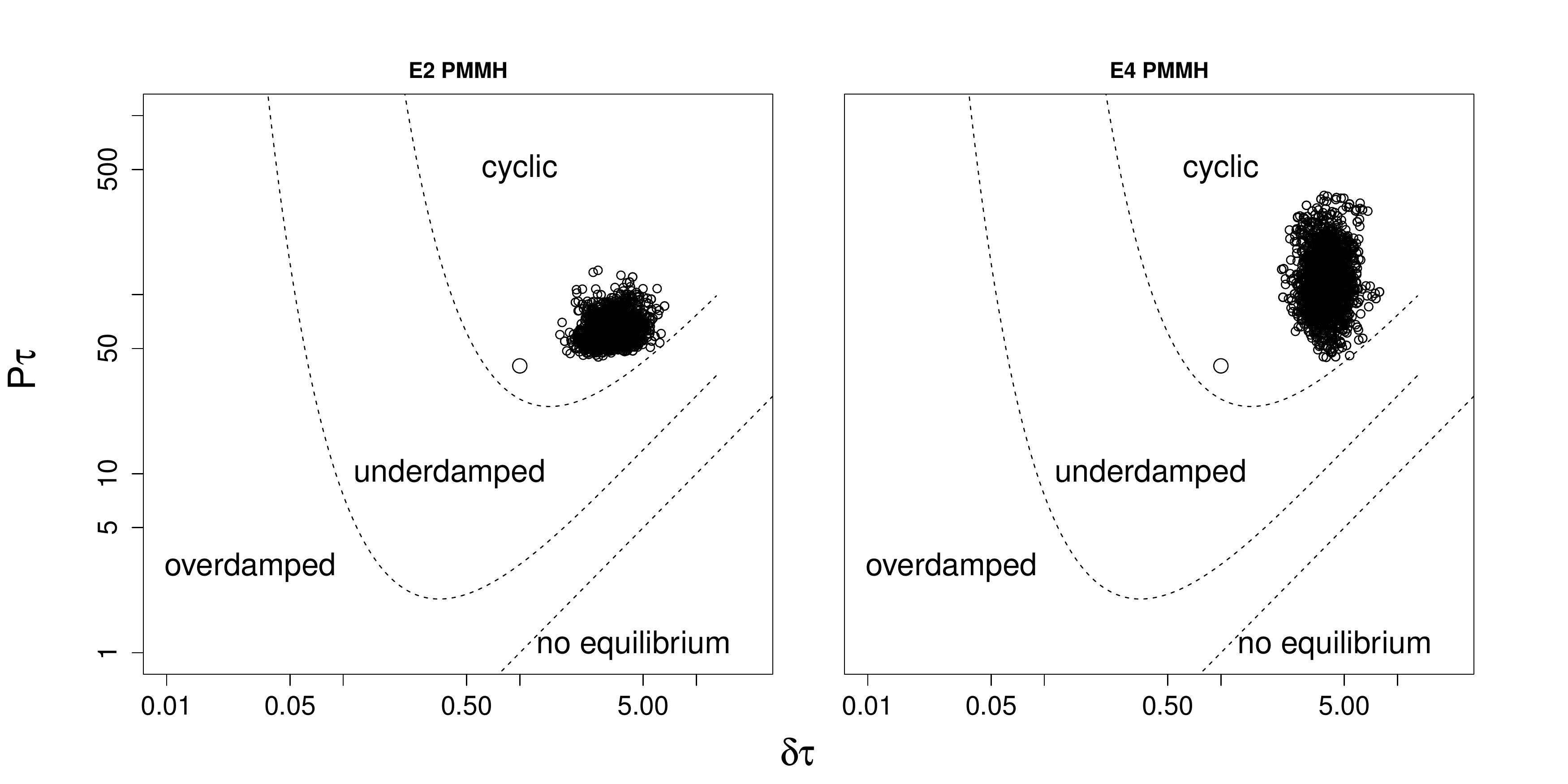}
\caption{Stability plots for datasets E2 and E4 using PMMH with log Student's t observational error.}
\label{fig:studBlow}
\end{figure}

To understand what happened, we have run a filtering operation using dataset E2, $10^4$ particles and parameters equal to the posterior mean given by SL and PMMH. Figure \ref{fig:ESS} shows the dynamics of the Effective Sample Size (ESS) using either parameter set. From the top plot we see that the ESS drops to practically zero around the 25th, 95th and 250th observation, if SL estimates are used. On the other hand, PMMH gives much higher estimates of $\sigma_p$ and $\sigma_d$ and this keeps the ESS from dropping to zero on those occasions. This suggests that few idiosyncrasies or outliers in datasets E2 and E4 might be pushing PMMH toward the underdamped region. This is supported by the fact that, if PMMH is run using a log Student's t-distribution for the observational process
$$
\frac{\log{y_t} - \log{n_t}}{\sigma_o} \sim \text{Student}(\nu = 2),  
$$   
the resulting posterior estimates for E2 and E4 lay strictly inside the cyclic region, as shown in Figure \ref{fig:studBlow}. We comment on these results in Section \ref{sec: discussion}.  

%While SL essentially ignores these features, the particle filter detects them.    is not adequate for the purpose of forecasting the short term dynamics shown by the observed datasets (E2 and E4 in particular). This doesn't affect SL, but it is an obstacle for particle-filter-based methods, which do try to sequentially predict the blowfly population density (see Appendix \ref{app:SIR}). Hence the PMMH algortihm struggles to find sets of parameter values that give a good predictive performance, and it moves instead to areas of the parameter space characterized by stable dynamics and high noise levels. This is consistent with the fact that PMMH gives high estimates for $\sigma_d^2$ and $\sigma_p^2$ for datasets E2 and E4, as shown in Table \ref{tab:blowEstim}.

\subsection{Example 3: Cholera epidemics in the Bay of Bengal} \label{subsec: Cholera}

\begin{figure}
\centering
\includegraphics[scale = 0.24]{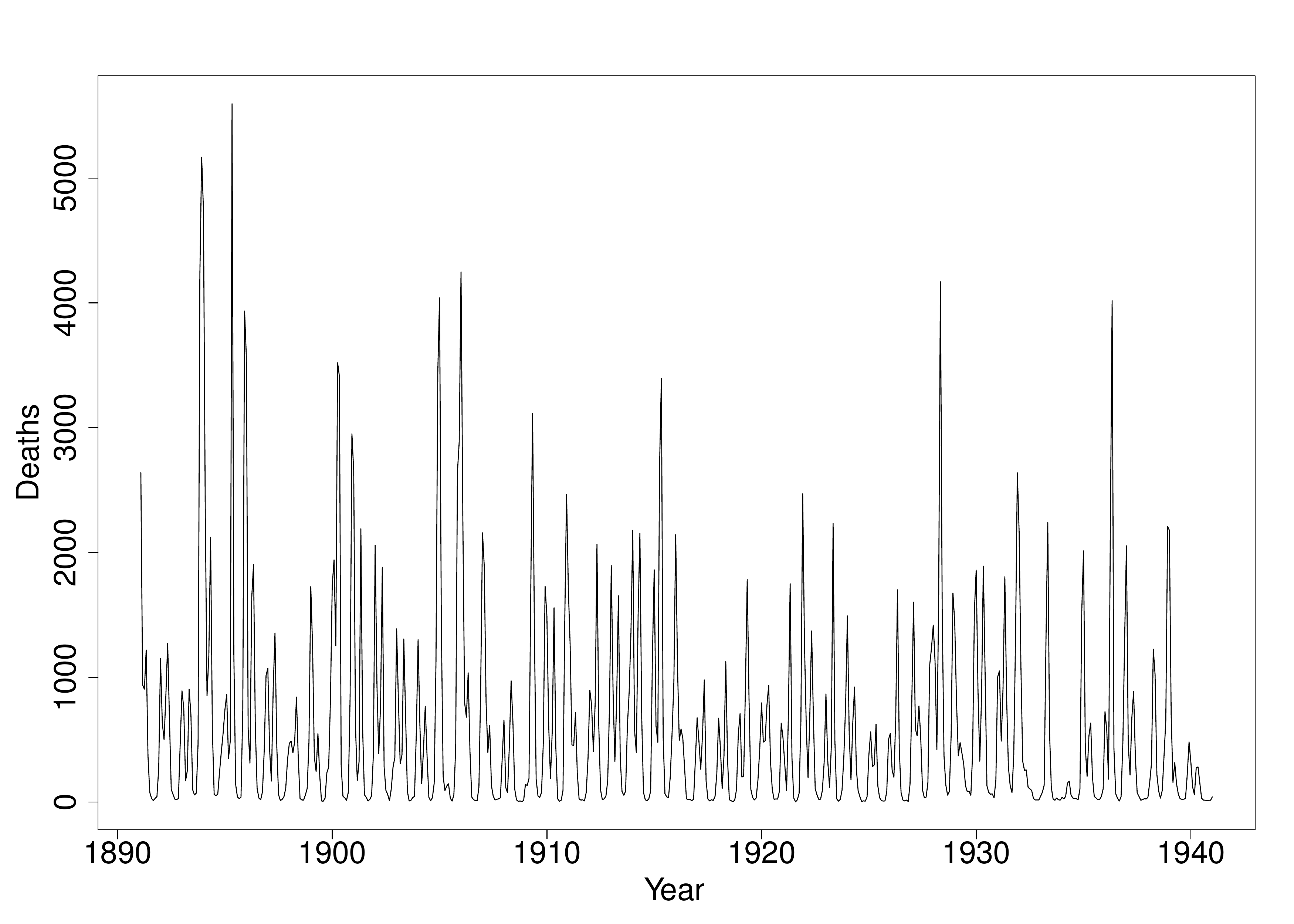}
\caption{Cholera-related monthly death count in the Dacca district between 1891 and 1941.}
\label{fig:dacca_plot}
\end{figure}

As a final example we consider a modified version of the Susceptible-Infected-Recovered-Susceptible (SIRS) model used by \cite{king2008inapparent} to explain cholera epidemics in the regions north of the Bay of Bengal. The dataset considered here corresponds to cholera-related mortality records in the former Dacca district of British East Indian province of Bengal, which is available within the \emph{pomp} R-package \citep{pomppack}. The data, depicted in Figure \ref{fig:dacca_plot}, consists of monthly deaths counts occurring between 1891 and 1941. See \cite{king2008inapparent} for additional details regarding the data.

\subsubsection{The model}

The model proposed by \cite{king2008inapparent} is composed of several classes, all of which are completely unobserved apart from the infected class, which is observed indirectly through the deaths count. In \cite{king2008inapparent} the model was represented by a system of differential equations, which was solved numerically using a Euler-Maruyama scheme. The main issue with their formulation is that the positivity of the states is not guaranteed. To address this problem, we propose an alternative model formulation, to be justified later, which results in the following system of difference equations
\begin{flalign*}
s_{t+1} &=  s_t - s_t^o + \frac{r_{kt}^ok \epsilon}{k \epsilon + \delta} + \frac{y_t^o \rho}{\rho + \delta} + b_{t+1}, \nonumber \\ 
i_{t+1} &=  i_t - i_t^o + c \frac{s_t^o\lambda_t}{\lambda_t + \delta},  \nonumber \\ 
y_{t+1} &=  y_t - y_t^o + (1-c) \frac{s_t^o\lambda_t}{\lambda_t + \delta}, \nonumber \\
r_{1t+1} &=  r_{1t} - r_{1t}^o + \frac{i_t^o\gamma}{m+\gamma+\delta},  \nonumber \\
r_{it+1} &=  r_{it} - r_{it}^o + \frac{r_{i-1t}^ok \epsilon}{k \epsilon + \delta}, \;\;\;\; \text{for} \;\; i = 2, \dots, k, \nonumber
\end{flalign*}
where
\begin{flalign} \label{eq:cholModel}
b_{t+1} & =  p_{t+1} - p_t + \frac{s_t^o \delta}{\lambda_t + \delta} + \frac{i_t^o \delta}{m + \gamma + \delta} \nonumber \\ 
        & +  \frac{y_t^o\delta}{\rho + \delta} + \sum_{i=1}^k  \frac{r_{it}^o\delta}{k \epsilon + \delta}, \nonumber  \\
s_{t}^o & =  s_t (1 - e^{-(\lambda_t + \delta) \Delta t}), \nonumber \\ 
i_{t}^o & =  i_t (1 - e^{-(m + \gamma + \delta) \Delta t}), \nonumber  \\ 
y_{t}^o & =  y_t (1 - e^{-(\rho + \delta) \Delta t}), \nonumber \\
r_{it}^o & =  r_{it} (1 - e^{-(k \epsilon + \delta) \Delta t}), \;\;\;\; \text{for} \;\;\ i = 1, \dots, k. \nonumber  \\
\end{flalign} 
Here $b_{t+1}$ represents the number of births between time $t$ and $t+1$, while $p_t$ is the total population of the Dacca district at time $t$, characterized by constant birth-death rate $\delta$. Susceptible individuals $s$ are infected by cholera at time-varying rate $\lambda_t$, which will be explained in detail later. Parameter $c$ determines the fraction of infected individuals that will undergo a full blown infection, represented by class $i$, rather than an asymptomatic infection, represented by class $y$. Individuals in $i$ suffer from an excess death rate $m$ and transition to the first Recovered class $r_1$ with rate $\gamma$. On the other hand, individuals in $y$ have the same death rate as susceptible individuals and do not acquire any long term immunity, as they rejoin the $s$ class directly at rate $\rho$. The duration of immunity is gamma distributed, with mean $1 / \epsilon$ and variance $k / \epsilon^2$.  

The rationale behind our discretized model needs to be clarified. Consider, for instance, $y_t$. To obtain $y_{t+1}$ we model inputs and outputs involving $y$ in turn, rather than simultaneously. Firstly, we obtain the number of individuals, $y^o_t$, leaving the  asymptomatic infected class by solving
$$
dy_s = -(\rho + \delta) y_s ds,
$$
between $t$ and $t+1$. The resulting solution is an exponential decay, which ensures the positivity of $y_{t+1}$. Then $y^o_t$ is divided between $b_{t+1}$ and $s_{t+1}$, with proportions determined by the output rates $\delta$ and $\rho$. This solution preserves the positivity of all classes and mass-balance, both of which are essential for a realistic model. In addition, our formulation becomes equivalent to the Euler-Maruyama scheme of \cite{king2008inapparent}, as $\Delta t \rightarrow 0$. 
 
The force of infection $\lambda_t$ is given by
\begin{equation} \label{eq:force}
\lambda_t = \omega_t + e^{\beta t}\beta_t \frac{i_{t}}{p_t}\frac{\Delta w}{\Delta t},
\end{equation}
where $\Delta w \sim \Gamma(\Delta t / \sigma^2, 1 / \sigma^2)$, so that $\Delta w / \Delta t$ represents multiplicative gamma noise with unit mean and variance equal to $\sigma^2$. We preferred this choice to the additive Gaussian noise originally used by \cite{king2008inapparent}, because the multiplicative version assures the positivity of $\lambda_t$. 

In (\ref{eq:force}), $\omega_t$ and $\beta_t$ represent respectively the environmental and human feedback components of the force of infection
$$
\omega_t = \exp{ \bigg ( \sum_{i = 1}^6 \omega_i g_i(t) \bigg ) }, 
$$
$$
\beta_t = \exp{ \bigg ( \sum_{i = 1}^6 \beta_i g_i(t) \bigg ) }, 
$$
where $g_i(t)$, for $i = 1, \dots, 6$, are a periodic B-spline basis.  Parameter $\beta$ is the long term trend in human-to-human transmission. 

The observed number of deaths registered during the $n$-th month, is assumed to follow a negative binomial distribution
$$
e_n \sim \text{NB} \bigg ( q_n, \frac{1}{\tau^2} \bigg ),
$$
with mean $q_n$ and variance $q_n + q_n^2 / \tau^2$, where $q_n$ is the accumulated number of cholera-related deaths between the previous and the current month
$$
q_n = \sum_{s = t_{n-1}}^{t_n} m i_s.
$$
In the original model $e_n$ was normally distributed around $q_n$, but that choice often produces negative death counts when the model is simulated. See \cite{king2008inapparent} for further model details.

\subsubsection{Set-up and results using the Dacca dataset}

Similarly to \cite{king2008inapparent}, we do not fit the full model, but we consider:
\begin{itemize}
\item a seasonal model where the $y$ class is not included ($c = 1$);
\item a two-path model were the environmental force of infection is constant ($\omega_s(t) = \omega_s$); 
\item a basic SIRS model where $c = 1$, $\omega_s(t) = \omega_s$ and $\beta_s(t) = \beta_s$. 
\end{itemize}
We fitted each model to the Dacca dataset using SL and PMMH. For both methods we used $1.4\times10^6$ MCMC iterations, the first half of which was discarded as burn-in period, and $2000$ simulations to estimate the (synthetic) likelihood at each step. We used uniform or diffuse priors for all parameters. We report them, together with the 26 summary statistics used by SL, in the Supplementary Material.

%The convergence performance of the chains was mixed. With the full model the mixing was quite poor under either method, with some parameters appearing to be unidentifiable under SL. On the other hand, most of the parameters of the seasonal model seem to be well identified using either method, with some important exceptions to be discussed later. 

Table \ref{tab:cholPosterior} reports the estimated Akaike Information Criterion (AIC) and the time needed to obtain a single estimate of $p(\bm y^0| \bm \theta)$ or $p(\bm s^0| \bm \theta)$, on a single core of a 3.60GHz i7-3820 CPU, for each model and method. SL and PMMH agree in selecting the seasonal reservoir model, while the two paths mechanisms does not improve the fit enough, relatively to the SIRS model, to justify the additional complexity. This is in contrast with the results of \cite{king2008inapparent}, whose second-order AIC estimate was lower for the two paths than for the SIRS model. 

Almost all the marginal posterior variances were higher when SL was used, with a median increase equal to 7.2, 2.6 and 2.2 for the seasonal, two paths and SIRS model, respectively. The variance increases were highest for the seasonal coefficients, $\omega_{1:6}$, of the force of infection, which suggest that the amount of information lost through the use of summary statistics is sizeable.     

\begin{table}[ht] 
\begin{center}
\begin{tabular}{rlll}
  \hline
 Method & Seasonal & Two Paths & SIRS   \\ 
  \hline
 $\text{AIC}_{\text{SL}}$ & -38.4 & -31.6 & -34.6 \\ 
 $\text{AIC}_{\text{PMMH}}$ & 7458 & 7532.6 & 7528.2   \\ 
 $\text{CPU}_{\text{SL}}$ & 10 & 10.3 & 9.8   \\ 
 $\text{CPU}_{\text{PMMH}}$ & 9.6 & 10.1 & 9.4   \\ 
  \hline
\end{tabular}
\end{center}
\caption{Estimated AICs and CPU times (sec) for each model, using SL and PMMH.}
\label{tab:cholPosterior}
\end{table}

\begin{figure}[h]
\centering
\includegraphics[scale = 0.30]{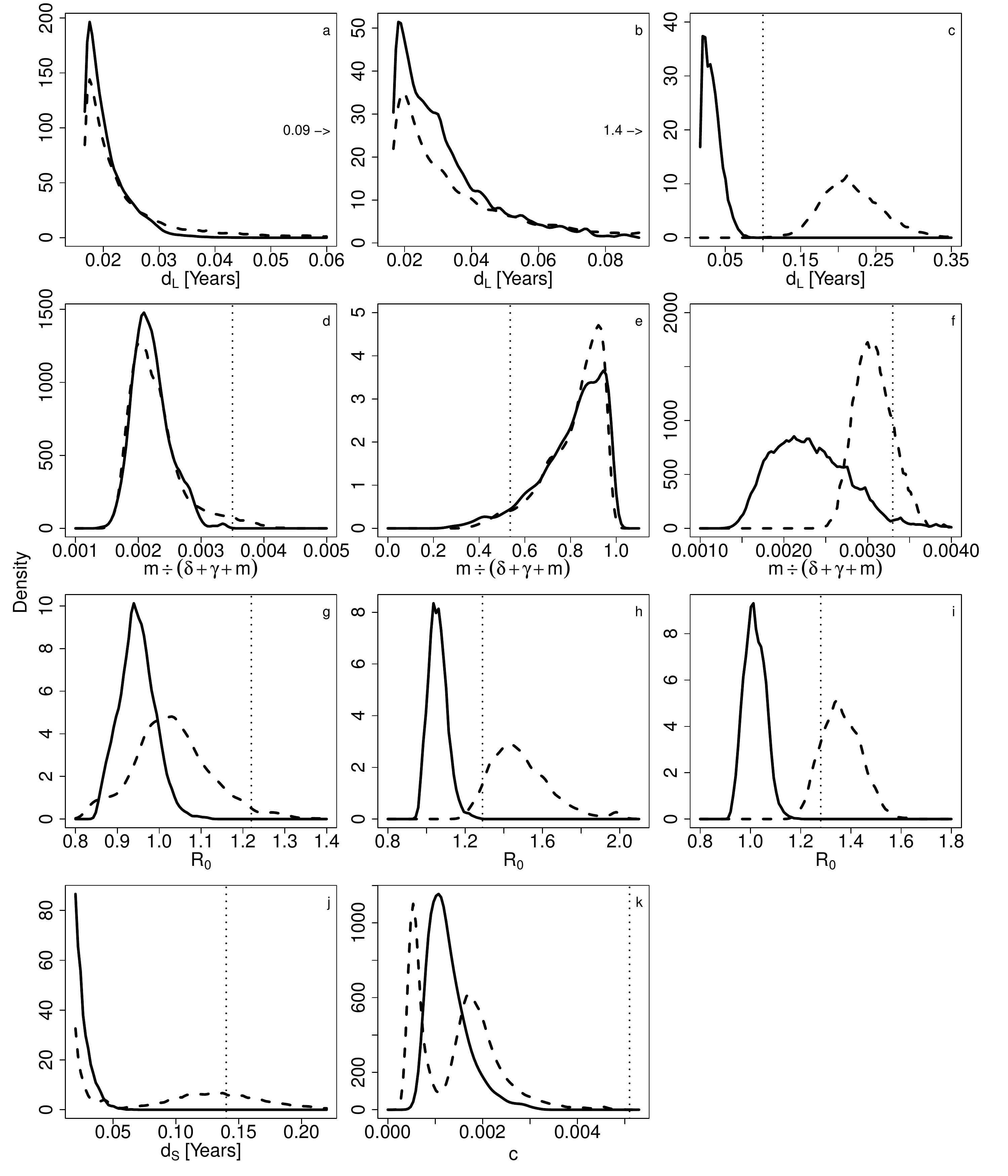}
\caption{Posterior marginal distributions from PMMH (solid) and SL (dashed). The estimates of \cite{king2008inapparent} correspond to the vertical dotted lines, substituted by annotations when out of range. The first three rows contain the marginals of immunity duration after full-blow infections, fatality and basic reproductive number for the seasonal (a, d, g), two paths (b, e, h) and SIRS (c, f, i) model. The last row shows the marginals of immunity duration after mild infections (j) and of the fraction of severe infections (k) for the two paths model. }
\label{fig:chol_marg_post}
\end{figure} 

One important hypothesis examined by \cite{king2008inapparent} was that the mean duration of immunity, $d_L := 1 / \epsilon$, might be much shorter than previously thought. Our analysis partially supports this conclusion, as shown by Figure \ref{fig:chol_marg_post}. The plots in the top row show the marginal densities of $d_L$ under each model. Under the seasonal model, most of the posterior mass lies close to the lower prior boundary, corresponding to unrealistically low periods of immunity (shorter than one week). The posterior given by SL under the SIRS model is slightly less extreme, but it still suggests period of immunity of one to three months, which is much shorter than the 3 to 10 years time-scale suggested by several sources \citep{cash1974response, glass1982endemic, koelle2005refractory}. One surprising result is that, under the two paths model, $d_L$ is still estimated to be lower than one month. This is in contrast with the results of \cite{king2008inapparent}, who estimates $d_L$ to be around $1.4$ years, under the same model and dataset. The mean duration of immunity after mild infections $d_S = 1 / \rho$ is estimated to be shorter than three weeks under PMMH, while SL seems to have lost information regarding $d_S$, as the corresponding marginal posterior is bimodal and highly dispersed. 

Figure \ref{fig:chol_marg_post} shows also the marginal distributions of the cholera-related death probability $f = m / (\delta + \gamma + m)$. Under the seasonal and the SIRS models our estimates roughly agree with those of \cite{king2008inapparent}, but our fatality estimate is much higher than theirs when asymptomatic infections are included in the model. Similarly to \cite{king2008inapparent}, we estimate the fraction of infection that are symptomatic to be very low under the two path model.

\begin{figure}
\centering
\includegraphics[scale = 0.25]{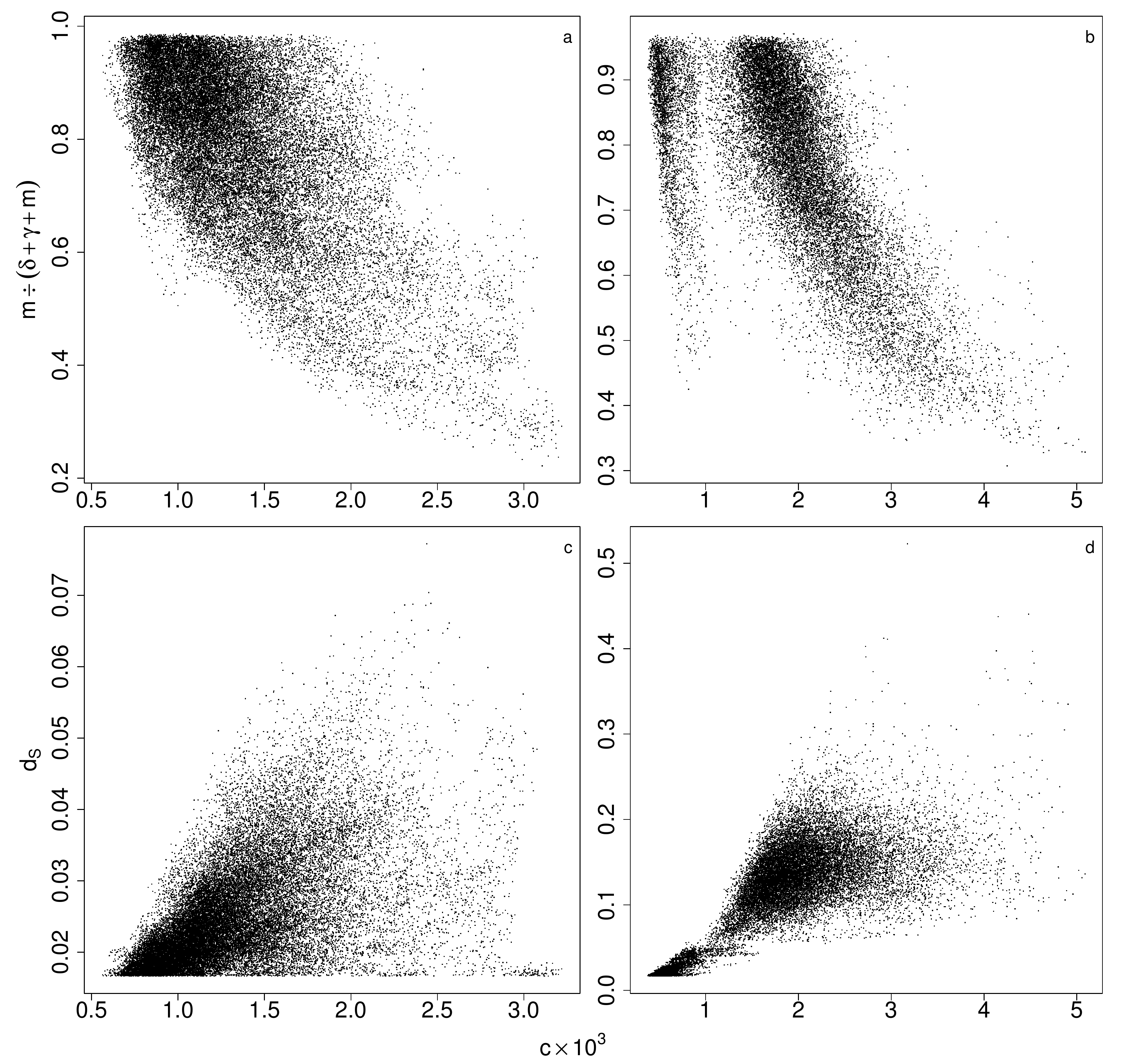}
\caption{Joint posterior samples for fraction of symptomatic infections vs fatality and duration of short term immunity under PMMH (a, c) and SL (b, d). }
\label{fig:chol_joint_post}
\end{figure}

Our results suggest that including asymptomatic infections does not improve the fit and does not provide more realistic estimates of immunity duration, following full-blown infections. In addition, this model is difficult to identify, because there is a trade-off between parameters $c$, $d_S$ and $m$, which is captured by Figure \ref{fig:chol_joint_post}. The correlations observed in the PMMH joint posterior sample are explained by the fact that an increase in the fraction of individuals with full infection can be compensated by decreasing their mortality rate or by increasing the duration of long short term immunity (thus delaying individuals with mild infection from rejoining the susceptible). Under SL this identifiability issue is more severe, and the corresponding posteriors are bimodal and more dispersed.

Another question addressed by \cite{king2008inapparent} is the relative importance of the environmental reservoir and of the human habitat for \emph{V.Cholerae} persistence. They found that the basic reproductive number, $\text{R}_0$, which quantifies the strength of human-to-human transmission, was consistently low (around 1.5) across model and geographic area. Figure \ref{fig:chol_marg_post} shows that our estimates of $\text{R}_0$ are very low under all models and methods, thus supporting the hypothesis that humans might be only a marginal habitat for \emph{V.Cholerae}. 

%In \cite{king2008inapparent} the modelobtained from model fitting is that 
%In the first place the estimates obtained by \cite{king2008inapparent} for the mean duration of immunity $1/\ \epsilon$ are much shorter than previ 

\section{Discussion} \label{sec: discussion}

We have described some of the difficulties that can be encountered when working with highly non-linear dynamical models, and we have shown how these issues influence the performance of some popular inferential approaches. In particular, in Section \ref{subsec: multi} we have provided strong experimental evidence suggesting that, when the dynamics of the system are chaotic or near-chaotic, the likelihood function becomes increasingly multimodal as the process noise is reduced. While this directly undermines the performance of state space methods aiming at estimating the full likelihood, as in PMMH, or its derivatives, as in IF, approaches based on information reduction are less affected. This has practical implications because, in an applied setting, it is generally not known whether the best fitting parameters lay in an area of the parameter space where the stochasticity is too low for state space methods to work adequately. Hence the ability of approaches based on information reduction to smooth the likelihood function, brought about by focusing on features of the data that are phase-independent, is appealing.

The blowflies example in Section \ref{subsec: blow}, highlights the robustness of information reduction methods from a different perspective. Indeed, careless application of PMMH would have classified the dynamics of the system as nearly-underdamped under two of Nicholson's datasets, with the corresponding simulations from the model being clearly inconsistent with the data (see Figure \ref{fig:dataBlow}). On the contrary, SL reliably classifies the dynamics as cyclic. In this example using a fat-tailed observation density mitigated the problem, but we argue that these results have deeper practical implications. Model \ref{eq: blowModel} has sufficient flexibility to reproduce the main features (quantified by the summary statistics) of Nicholson's datasets, as demonstrated by Figure \ref{fig:dataBlow}. On the other hand, the model struggles to explain certain nuances of Nicholson's datasets, and this is detected by the particle filter, but overlooked by SL. This suggests that, in situations in which the model has a clear scientific interpretation, but lacks the ability to explain the observed dynamics in all their complexity, focusing on some salient features of the data might be a reasonable approach. Conversely, if the model is believed to be an accurate description of the system under study, or if it is meant to be used for the purpose of state estimation or forecasting, then it is compelling to fit it using the full data.  

Another lesson learned from the blowflies example is that, for particle-filtering-based methods to work properly, a good initialization is often indispensable. This is because these methods are generally based on some form of importance sampling, hence when the initial estimates are far from the best fitting parameters most of the importance weights go to zero (particle depletion). In this context, methods based on information reduction can be useful, because they are robust to bad initializations. Methods that can provide reliable initial estimates, to be fed to more accurate but less robust methods, are of high practical value, but often under-represented in the literature. Exceptions are \cite{lavine2013} who, in the context of pertussis epidemics, use SL to initialize a IF algorithm and \cite{owen2014likelihood}, who proposes to initialize PMMH using the output of preliminary ABC runs.

One recurrent theme in our examples is that reducing the data to a set of summary statistics generally entails a loss of accuracy in parameter estimation. This is particularly clear in Section \ref{subsec: simpleMaps}, where SL and ABC are consistently outperformed by PMMH and IF in terms of MSEs. Mild losses of accuracy are often acceptable when parameter estimation is not the main focus of analysis, but the aim is, for example, to determine whether the dynamics of the system are stable or oscillatory, as in the blowflies example. On the other hand, when dealing with models that are weakly identified even under the full data, as in Section \ref{subsec: Cholera}, any further loss of information can lead to unreliable estimates. Hence, an important drawback of information reduction methods is that, in the absence of a benchmark, quantifying inferential inaccuracies require running simulation studies, which can be prohibitively expensive for complex models, such as those presented in Section \ref{subsec: Cholera}. While in all the examples presented in this study one or more benchmarks were available, this not always the case. 

All the methods described in this work, with the exception of Parameter Cascading, are computationally intensive. In particular, obtaining pointwise estimates of $p(\bm y^0_{1:T}|\bm \theta)$ or $\nabla p(\bm y^0_{1:T}|\bm \theta)$ requires $MT$ simulations, where $M$ is the number of particles, from $p(\bm n_t|\bm n_{t-1},\theta)$ under SIR and IF respectively. Similarly SL uses $N$ simulations from $p(\bm y_{1:T}|\bm \theta)$ to estimate $p(\bm s^{0}|\bm \theta)$. Within PMMH and the MCMC implementation of SL, this price has to be paid at each iteration and the efficiency of the sampler will depend on the trade-off between the variance of likelihood estimates and the number of simulations used to obtain them \citep{sherlock2014efficiency}. Similar considerations hold for IF, but the optimizer generally needs much fewer iterations to reach convergence. On the other hand IF does not directly provide parameter uncertainty estimates, which have to be obtained through an expensive likelihood profiling procedure (see \cite{ionides2006inference}). On first sight ABC samplers seem more efficient than the above approaches, because they target $p(\bm \theta|\bm s^0)$ directly, by simulating a single statistics vector at the time. However, ABC samplers generally have a very low acceptance rate, because the latter increases with the tolerance $\epsilon$, while their accuracy is inversely proportional to it. 

These computational issues are aggravated by the curse of dimensionality. In particular, the number of particles in a particle filter need to increase super-exponentially with the number of hidden states, in order to avoid particle-depletion \citep{snyder2008obstacles}. This result applies directly to PMMH and IF. Analogously, the computational cost of method based on information reduction typically increases with the number of summary statistics used ($d$). In ABC methods, the MSE of the posterior moments estimate decreases at rate $O(e^{-4/d+5})$, due to the non-parametric approximation used by such methods \citep{blum2010approximate}. SL scales better with $d$, because it requires a number of simulations sufficient to estimate the $O(d^2)$ entries of $\bm \Sigma_{\bm \theta}$. However, its Gaussian assumption might hold only approximately.

Summary statistics selection is, in our opinion, an open problem, as many approaches proposed in the literature require the user to specify an initial set of summary statistics which can then be refined upon (see for example \cite{blum2013comparative}, \cite{fearnhead2012constructing} or \cite{nunes2010optimal}). While some fairly general approaches exist \citep{drovandi2013bayesian}, finding a set of initial statistics under which the model is identifiable is, at the time of writing, a time consuming, problem dependent and largely non-automated process. In the context of models with several hidden states, devising summary statistics is particularly difficult, because these have to capture the relation between all the states, while being based only on (noisy proxies of) a subset of them. The two-path cholera model is a perfect example of this problem: out of seven state variables only one, the number of infected, is observed with noise.

Taken together our results lead us to some very practical conclusions. When faced with a real non-linear dynamic system for which good models are available, one should ideally use a state space method for final parameter estimation, combined with a minimum tuning information reduction approach for exploration of alternative model structures, initialization and checking of conclusions. Using state space methods alone may bias conclusions towards noise driven stable dynamics, while using information reduction alone may lead to inference that is less precise than it could be. If the model is only attempting to explain some features of the system, and not every detail of the data then information reduction is probably essential. 

\section*{Acknowledgements}
This work has been partly funded by the EPSRC grant EP/I000917/1 and EP/K005251/1. The authors would like to thank Aaron King and Ed Ionides for useful discussion and Chris Jennison for commenting on earlier versions of this work and the suggestion that led to the discretized Ricker model.

\bibliographystyle{imsart-nameyear}
\bibliography{biblio}

\pagebreak
\begin{center}
\textbf{\large Supplementary Material}
\end{center}
%%%%%%%%%% Merge with supplemental materials %%%%%%%%%%
%%%%%%%%%% Prefix a "S" to all equations, figures, tables and reset the counter %%%%%%%%%%
\setcounter{equation}{0}
\setcounter{section}{0}
\setcounter{figure}{0}
\setcounter{table}{0}
\setcounter{page}{1}
\makeatletter
\renewcommand{\theequation}{S\arabic{equation}}
\renewcommand{\thefigure}{S\arabic{figure}}
\renewcommand{\bibnumfmt}[1]{[S#1]}
\renewcommand{\citenumfont}[1]{S#1}

\section{Discretized SSM} \label{app:discrApp}

The likelihood of a simple SSM can be written in the following form
$$
p(\bm y_{1:T}|\bm \theta) =  p(\bm y_1|\bm \theta)\prod_{t = 2}^Tp(\bm y_t|\bm y_{1:t-1},\bm \theta),
$$
and, if $m$ is the number of discrete levels of the hidden state, then
\begin{equation*} \label{eq: discreteFilter}
p(\bm y_1|\bm \theta) = \sum_{i = 1}^m p(\bm y_1|\bm n^i_1,\bm \theta)p(\bm n^i_1|\bm \theta),
\end{equation*}
and
\begin{flalign*}
& p(\bm y_t|\bm  y_{1:t-1},\bm \theta) =  \sum_{i = 1}^m p(\bm y_t|\bm n^i_t,\bm \theta)p(\bm n^i_t|\bm y_{1:t-1},\bm \theta)  \nonumber \\ 
& = \sum_{i = 1}^m p(\bm y_t|\bm n^i_t,\bm \theta) 
         \sum_{j = 1}^m p(\bm n^i_t|\bm n^j_{t-1},\bm \theta)p(\bm n^j_{t-1}|\bm y_{1:t-1}, \bm\theta), \nonumber  \\
\end{flalign*}
where
\begin{flalign*}
& p(\bm n^j_{t-1}|\bm y_{1:t-1},\bm  \theta) = \frac{\sum_{k=1}^m p(\bm y_{1:t-1},\bm n^j_{t-1}, \bm n^k_{t-2}|\bm \theta)}{p(\bm y_{1:t-1}|\bm  \theta)}  \nonumber \\ 
& =   p(\bm y_{t-1}|\bm n^j_{t-1}, \bm  \theta) \sum_{k=1}^m p(\bm n^j_{t-1}| \bm n^k_{t-2}, \bm \theta)p(\bm n^k_{t-2}|\bm y_{1:t-2}, \bm \theta) \nonumber \\
& \times \frac{p(\bm y_{1:t-2}| \bm \theta)}{p(\bm y_{1:t-1}| \bm \theta)} \nonumber \\
& = p(\bm y_{t-1}|\bm n^j_{t-1},\bm \theta) \sum_{k=1}^m p(\bm n^j_{t-1}|\bm  n^k_{t-2}, \bm \theta) \frac{ p(\bm n^k_{t-2}|\bm y_{1:t-2},\bm \theta)}{p(\bm y_{t-1}|\bm y_{1:t-2},\bm \theta)}. \nonumber \\
\end{flalign*}
These formulas can be used to calculate the likelihood of a discrete SSM exactly.

\section{Computational details} \label{app:comDetails}

To fit the models described in this work we used the \emph{synlik} \citep{synlikPack}, \emph{EasyABC} \citep{jabot2013easyabc} and \emph{pomp} \citep{pomppack}  R-packages. The first two provide implementations of SL and ABC respectively, while we used \emph{pomp} to run the IF and PMMH algorithms.

\subsection{Simple maps} \label{app:simpleApp}

The data was simulated using the following parameter values:
\begin{itemize}
\item Generalized Ricker: $r = 44.7$, $\theta = 1$, $\sigma = 0.3$, $\phi = 10$.
\item Pennycuick: $r = 58$, $a = 0.1$, $\sigma = 0.3$, $\phi = 1$.
\item Maynard-Smith: $r = 18$, $b = 6$, $\sigma = 0.4$, $\phi = 24$.
\item Varley: $r = 15$, $b = 5.5$, $c = 1$, $\sigma = 0.45$, $\phi = 20$.
\end{itemize}

For SL and ABC-MCMC we used the set of 13 summary statistics proposed by \cite{wood2010}:
\begin{itemize}
\item the autocovariances of the path $y_{1:T}$ up to lag 5;
\item the mean population $\bar{y}$;
\item the number of zeros observed;
\item the coefficients of the regression
$$
y_{t+1}^{0.3}=\beta_1y_t^{0.3}+\beta_2y_t^{0.6}+z_t;
$$
\item the coefficients of a cubic regression of the ordered differences $y_t-y_{t-1}$ on their observed values.
\end{itemize} 

Tables \ref{tab:rickerPrior} to \ref{tab:varleyPrior} contain the limits of the uniform priors (or box constraints under IF) and initial values used for each model and parameter.

\begin{table}[h]
\begin{center}
\begin{tabular}{rrrr}
  \hline
 & Initial & Lower & Upper \\ 
  \hline
r & 2.80 & 2.00 & 5.00 \\ 
  $\sigma$ & -2.30 & -3.00 & -0.22 \\ 
  $\phi$ & 1.79 & 1.61 & 3.00 \\ 
   \hline
\end{tabular}
\end{center}
\caption{Prior boundaries for Ricker} \label{tab:rickerPrior}
\end{table} 

\begin{table}[h]
\begin{center}
\begin{tabular}{rrrr}
  \hline
 & Initial & Lower & Upper \\ 
  \hline
r & 2.80 & 2.00 & 5.00 \\ 
  $\theta$ & 0.41 & -0.69 & 0.41 \\ 
  $\sigma$ & -2.30 & -3.00 & -0.22 \\ 
  $\phi$ & 1.79 & 1.61 & 3.00 \\ 
   \hline
\end{tabular}
\end{center}
\caption{Prior boundaries for Generalized Ricker} 
\end{table}

\begin{table}[h]
\begin{center}
\begin{tabular}{rrrr}
  \hline
 & Initial & Lower & Upper \\ 
  \hline
r & 3.69 & 2.50 & 5.00 \\ 
  a & -1.20 & -4.61 & -0.69 \\ 
  $\sigma$ & -0.69 & -3.00 & -0.22 \\ 
   \hline
\end{tabular}
\end{center}
\caption{Prior boundaries for Pennycuick} 
\end{table}

\begin{table}
\centering
\begin{tabular}{rrrr}
  \hline
 & Initial & Lower & Upper \\ 
  \hline
r & 2.30 & 1.50 & 4.00 \\ 
  b & 2.20 & 0.69 & 2.30 \\ 
  $\sigma$ & -0.69 & -3.00 & -0.22 \\ 
  $\phi$ & 2.64 & 2.30 & 3.56 \\ 
   \hline
\end{tabular}
\vspace*{1mm}
\caption{Prior boundaries for Maynard-Smith} 
\end{table}

\begin{table}[h!]
\begin{center}
\begin{tabular}{rrrr}
  \hline
 & Initial & Lower & Upper \\ 
  \hline
r & 2.30 & 1.50 & 4.00 \\ 
  b & 2.01 & 0.69 & 2.30 \\ 
  C & 0.69 & -2.30 & 0.69 \\ 
  $\sigma$ & -1.61 & -3.00 & -0.22 \\ 
  $\phi$ & 2.71 & 2.30 & 3.40 \\ 
   \hline
\end{tabular}
\end{center}
\caption{Prior boundaries for Varley} \label{tab:varleyPrior}
\end{table} 

Tables \ref{tab:rickerMSE} to \ref{tab:varleyMSE} contain the root median squared errors (MSE) and coverage frequencies for each parameter of the five models considered, using each method. The last row indicates which method achieved the lowest mean squared error, for each model parameter.

\begin{table}
\begin{center}
\begin{tabular}{rlll}
  \hline
 & r & $\sigma$ & $\phi$ \\ 
  \hline
SL & 0.11(0.9) & 0.34(0.92) & 0.05(0.88) \\ 
  SL\_R & 0.12(0.9) & 0.34(0.92) & 0.05(0.88) \\ 
  ABC & 0.14(0.96) & 0.2(1) & 0.04(1) \\ 
  IF & 0.11(-) & 0.28(-) & 0.03(-) \\ 
  PMMH & 0.1(1) & 0.21(1) & 0.02(1) \\ 
  Best & PMMH & ABC & PMM\vspace*{1mm}H \\ 
   \hline
\end{tabular}
\end{center}
\caption{RMSEs(coverage) for Ricker} \label{tab:rickerMSE}
\end{table}

\begin{table}
\begin{center}
\begin{tabular}{rllll}
  \hline
 & r & $\theta$ & $\sigma$ & $\phi$ \\ 
  \hline
SL & 0.24(0.92) & 0.06(0.98) & 0.4(0.86) & 0.17(0.96) \\ 
  SL\_R & 0.23(0.96) & 0.06(1) & 0.41(0.92) & 0.17(0.98) \\ 
  ABC & 0.16(0.98) & 0.04(1) & 0.16(1) & 0.13(1) \\ 
  IF & 0.13(-) & 0.03(-) & 0.3(-) & 0.1(-) \\ 
  PMMH & 0.12(0.94) & 0.03(1) & 0.23(0.98) & 0.11(0.98) \\ 
  Best & PMMH & IF & ABC & IF \\ 
   \hline
\end{tabular}
\end{center}
\caption{RMSEs(coverage) for Generalized Ricker} 
\end{table}

\begin{table}
\begin{center}
\begin{tabular}{rlll}
  \hline
 & r & a & $\sigma$ \\ 
  \hline
SL & 0.14(0.9) & 0.05(0.94) & 0.34(0.98) \\ 
  SL\_R & 0.15(0.9) & 0.04(0.94) & 0.34(1) \\ 
  ABC & 0.14(1) & 0.07(1) & 0.14(1) \\ 
  IF & 0.11(-) & 0.03(-) & 0.26(-) \\ 
  PMMH & 0.1(0.92) & 0.02(0.98) & 0.19(0.92) \\
  Best & PMMH & PMMH & ABC \\ 
   \hline
\end{tabular}
\end{center}
\caption{RMSEs(coverage) for Pennycuick} 
\end{table}

\begin{table}
\begin{center}
\begin{tabular}{rllll}
  \hline
 & r & b & $\sigma$ & $\phi$ \\ 
  \hline
SL & 0.13(0.92) & 0.25(1) & 0.43(0.88) & 0.24(1) \\ 
  SL\_R & 0.13(0.94) & 0.2(1) & 0.44(0.88) & 0.22(1) \\ 
  ABC & 0.11(1) & 0.25(1) & 0.17(1) & 0.23(1) \\ 
  IF & 0.12(-) & 0.45(-) & 0.29(-) & 0.48(-) \\ 
  PMMH & 0.09(0.98) & 0.13(1) & 0.23(0.96) & 0.12(1) \\ 
  Best & PMMH & PMMH & ABC & PMMH \\ 
   \hline
\end{tabular}
\end{center}
\caption{RMSEs(coverage) for Hassell} 
\end{table}

\begin{table}
\begin{center}
\begin{tabular}{rllll}
  \hline
 & r & b & $\sigma$ & $\phi$ \\ 
  \hline
SL & 0.16(0.9) & 0.07(0.88) & 0.61(0.78) & 0.12(0.94) \\ 
  SL\_R & 0.15(0.96) & 0.06(0.9) & 0.67(0.92) & 0.1(1) \\ 
  ABC & 0.19(0.94) & 0.06(1) & 0.27(1) & 0.09(1) \\ 
  IF & 0.11(-) & 0.04(-) & 0.26(-) & 0.06(-) \\ 
  PMMH & 0.09(1) & 0.04(1) & 0.15(1) & 0.05(1) \\  
  Best & PMMH & PMMH & PMMH & PMMH \\ 
   \hline
\end{tabular}
\end{center}
\caption{RMSEs(coverage) for Maynard-Smith} 
\end{table}

\begin{table*}[ht]
\begin{center}
\begin{tabular}{rlllll}
  \hline
 & r & b & C & $\sigma$ & $\phi$ \\ 
  \hline
SL & 0.16(0.96) & 0.07(0.92) & 0.16(1) & 0.87(0.76) & 0.1(0.92) \\ 
  SL\_R & 0.16(0.98) & 0.07(0.96) & 0.17(1) & 0.8(0.88) & 0.11(0.94) \\ 
  ABC & 0.17(0.98) & 0.06(1) & 0.17(1) & 0.32(1) & 0.07(1) \\ 
  IF & 0.1(-) & 0.05(-) & 0.12(-) & 0.34(-) & 0.07(-) \\ 
  PMMH & 0.1(0.96) & 0.04(0.96) & 0.08(1) & 0.2(0.94) & 0.06(0.96) \\ 
  Best & IF & PMMH & PMMH & PMMH & PMMH \\ 
   \hline
\end{tabular}
\end{center}
\caption{RMSEs(coverage) for Varley}  \label{tab:varleyMSE}
\end{table*}

\subsection{Blowflies} \label{app: blowApp}

For this model we used the set of 16 summary statistics proposed by \cite{wood2010}:
\begin{itemize}
\item the autocovariances of the path $n_{1:T}$ up to lag 11;
\item the mean population $\bar{n}$;
\item the difference between mean and median population $\bar{n} - \tilde{m}$;
\item the number of zeros observed;
\item the coefficients of the regression 
$$
n_{t+1}=\beta_1n_t + \beta_2n_t^2 + \beta_3n_t^3 + \beta_4n_{t-6} + \beta_5n_{t-6}^2 + z_t;
$$
\item the coefficients of a cubic regression of the ordered differences $n_t-n_{t-1}$ on their observed values.
\item the number of turning points.
\end{itemize}

The priors used when fitting the simulated datasets are reported in Table \ref{tab:blowSimulPrior}.

\begin {table}
\begin{center}
    \begin{tabular}{| c | c | }
    \hline
    Parameter & Prior \\ \hline
    $\delta$ & $\text{Unif}(0.09, 0.4)$  \\ \hline
    $P$ & $\text{Unif}(3, 12)$ \\ \hline
    $N_0$ & $\text{Unif}(150, 800)$ \\ \hline
    $\sigma_p^2$ & $\text{Unif}(0.01, 1)$ \\ \hline
    $\tau$ & $\text{Unif}(5, 25)$ \\ \hline
    $\sigma_d^2$ & $\text{Unif}(0.01, 1)$ \\ \hline
    \end{tabular}
\end{center}
\caption{Priors used for the blowfly model in the simulated setting.}
\label{tab:blowSimulPrior}
\end{table}

Table \ref{tab:blowTruePrior} reports the priors used when fitting Nicholson's datasets. Notice that for $\tau$ we have used a non-uniform prior, based on information reported by \cite{gur1980nich} concerning biologically plausible values of this delay parameter.

\begin {table}
\begin{center}
    \begin{tabular}{| c | c | }
    \hline
    Parameter & Prior \\ \hline
    $\delta$ & $\text{Unif}(0.02, 1)$  \\ \hline
    $P$ & $\text{Unif}(3, 30)$ \\ \hline
    $N_0$ & $\text{Unif}(10, 1000)$ \\ \hline
    $\sigma_p^2$ & $\text{Unif}(0.01, 5)$ \\ \hline
    $\tau$ & $\text{Norm}(\mu = 14, \sigma = 5)$ \\ \hline
    $\sigma_d^2$ & $\text{Unif}(0.01, 5)$ \\ \hline
    \end{tabular}
\end{center}
\caption{Priors used for the blowfly model when fitting Nicholson's datasets.}
\label{tab:blowTruePrior}
\end{table}  

\subsection{Cholera in Dacca} \label{app:cholApp}

One thing to notice about model (5.3) is that cholera-related deaths
$$
D_t = \frac{ I^o_t m } { \gamma + \delta + m },
$$
are not offset by an equal number of births in the susceptible compartment $S_{t+1}$. Beside not making sense biologically, this would introduce a strong feedback mechanism during epidemics. To offset this downward bias on total population, we tilt the number of births at each step as follows
$$
B^*_{t+1} = B_{t+1} + \bar{D} \Delta t
$$
where $\bar{D}$ is the monthly average of the observed number of deaths during the whole period and $\Delta t$ is the time step used. $B^*_t$ is then used in place of $B_t$ in (5.3). With this choice the sum of the number individuals in each compartment does not match the official census, but we have verified that the mismatch is minimal. 

Let $d_t$ be the number of cholera-related deaths during the $t$-month, and define $r_t = d_t^{1/5}$. For SL we used the following set of 26 summary statistics: 
\begin{itemize}
\item the coefficients (intercept excluded) of the regression
\begin{flalign*}
r_{t} & =  \alpha_1 + \alpha_2 t + \sum_{i = 1}^4 \alpha_{3i} sin(\psi_i 2 \pi t) \nonumber \\ 
      & + \alpha_{4i} cos(\psi_i 2 \pi t) + z_t; \nonumber  \\
\end{flalign*}
where $\psi_1 = 0.12$, $\psi_2 = 1$, $\psi_3 = 2$, and $\psi_4 = 3$. 
Let $e_t$ be the $t$-residual of such regression;
\item the autocovariances of $e_{1:T}$ at lag 2, 6, and 11;
\item the mean $\bar{d}$ and variance $\text{Var}(d)$ of the number of deaths;
\item the scaled difference between mean and median number of deaths $(\bar{d} - \tilde{d}) / \text{Var}(d)$;
\item the coefficients of the auto-regression 
\begin{flalign*}
e_{t+1} & =  \beta_1e_t + \beta_2e_{t-2} + \beta_3e_{t-3}  \nonumber \\ 
        & + \beta_4e_{t-4} + \beta_{10}e_{t-10} + z_t; \nonumber  \\
\end{flalign*}
\item the coefficients of a cubic regression of the ordered differences $e_t-e_{t-1}$ on their observed values;
\item the number of turning points in $d_{1:T}$;
\item the median and inter-quartile range of $e_{1:T}$.
\end{itemize}
Table \ref{tab:cholPrior} reports the prior distributions used.
\begin {table}
\begin{center}
    \begin{tabular}{| c | c | }
    \hline
    Parameter & Prior \\ \hline
    $\gamma$ & $\text{Unif}(1, 365)$  \\ \hline
    $\epsilon$ & $\text{Unif}(0.1, 60)$ \\ \hline
    $c$ & $\text{Unif}(0, 1)$ \\ \hline
    $\rho$ & $\text{Unif}(1, 60)$ \\ \hline
    $m$ & $\text{Unif}(0, 140)$ \\ \hline
    $e^\beta$ & $\text{N}(0, 1000)$ \\ \hline
    $e^{\beta_1}, \dots, e^{\beta_6}$ & $\text{N}(0, 1000)$ \\ \hline
    $e^{\omega_1}, \dots, e^{\omega_6}$ & $\text{N}(0, 1000)$ \\ \hline
    $\sigma$ & $\text{Unif}(0, 1)$ \\ \hline
    $\tau$ & $\text{Unif}(0, 1)$ \\ \hline
    \end{tabular}
\end{center}
\caption{Priors used for the the Cholera model.}
\label{tab:cholPrior}
\end{table}  

Calculating the AICs reported in the main text was not straightforward, because the joint posterior distributions of the parameters are far from normal for each model, hence the posterior mean is inadequate as a point estimate. In addition, for both SL and PMMH the (synthetic) likelihood is estimated with noise, which makes finding good point estimates more difficult. To work around this issue, for each model and method, we restricted our attention to parameters corresponding to likelihood estimates above the 99th quantile and we have re-estimated the likelihood at each of those parameter values, using a $2 \times 10^4$ particles or simulations from the model. Given that these estimates had very low noise, we have used the parameter vector corresponding to highest likelihood estimate as a proxy for the MLE. Finally, we re-estimated the likelihood at the MLE using $5 \times 10^4$ simulations, and we have used it to estimate the AIC.

\end{document}